\documentclass[english,journal=jpcafh,manuscript=article]{achemso}
\usepackage[latin9]{inputenc}
\usepackage{color}
\usepackage{amsmath}
\usepackage{amssymb}
\usepackage{graphicx}

\makeatletter

\providecommand{\tabularnewline}{\\}

\usepackage{babel}
\author{Himanshu Chakraborty}
\author{Alok Shukla}
\email{chakraborty.himanshu@gmail.com,  shukla@phy.iitb.ac.in}
\affiliation{Department of Physics, Indian Institute of Technology Bombay, Powai, Mumbai 400076, INDIA}

\title[\texttt{achemso} demonstration]
{Pariser-Parr-Pople Model based Investigation of Ground and Low-Lying Excited States of Long Acenes}

\makeatother

\usepackage{babel}
\begin{document}
\begin{abstract}
{\normalsize Several years back Angliker }\emph{\normalsize et al}{\normalsize .
{[}Chem. Phys. Lett. }\textbf{\normalsize 1982}{\normalsize , }\emph{\normalsize 87}\textbf{\normalsize ,}{\normalsize {}
208{]} predicted nonacene to be the first linear acene with the triplet
state $1^{3}B_{2u}$ as the ground state, instead of the singlet $1^{1}A_{g}$
state. However, contrary to that prediction, in a recent experimental
work Tönshoff and Bettinger {[}}\textit{\emph{\normalsize Angew. Chem.
Int. Ed}}\emph{\normalsize .}{\normalsize {} }\textbf{\textit{\emph{\normalsize 2010}}}\textit{\emph{\normalsize ,
}}\textit{\normalsize 49}{\normalsize , 4125{]} demonstrated that
nonacene has a singlet ground state. Motivated by this experimental
finding, we decided to perform a systematic theoretical investigation
of the nature of the ground, and the low-lying excited states of long
acenes, with an emphasis on the singlet-triplet gap, starting from
naphthalene, all the way up to decacene. Methodology adopted in our
work is based upon Pariser-Parr-Pople model (PPP) Hamiltonian, along
with large-scale multi-reference singles-doubles configuration interaction
(MRSDCI) approach. Our results predict that even though the singlet-triplet
gap decreases with the increasing conjugation length, nevertheless,
it remains finite till decacene, thus providing no evidence of the
predicted singlet-triplet crossover. We also analyze the nature of
many-particle wavefunction of the correlated singlet ground state
and find that the longer acenes exhibit tendency towards a open-shell
singlet ground state. Moreover, when we compare the experimental absorption
spectra of octacene and nonacene with their calculated singlet and
triplet absorption spectra, we observe excellent agreement for the
singlet case. Hence, the optical absorption results also confirm the
singlet nature of the ground state for longer acenes. Calculated triplet
absorption spectra of acenes predict two well separated intense long-axis
polarized absorptions, as against one such peak observed for the singlet
case. This is an important prediction regarding the triplet optics
of acenes, which can be tested in future experiments on oriented samples. }{\normalsize \par}
\end{abstract}
\maketitle
\textcolor{black}{Keywords: oligoacenes, octacene, nonacene, optical
absorption spectrum, singlet-triplet gap, PPP model Hamiltonian, configuration-interaction
method.}

\section{{\normalsize Introduction}}

Polyacenes, which can be seen as linearly fused benzene rings, are
known for their well defined structures, and crystalline forms. \cite{clar1964polycyclic,clar1972aromatic,cooke1988polynuclear,harvey1991polycyclic,bjorseth1983handbook,Geerts1998}
Because of their small band gaps and high charge-carrier mobilities,
they find potential applications in novel opto-electronic devices
such as light-emitting diodes, and field effect transistors \emph{etc.,}
which make them experimentally and theoretically a very important
class of materials.\cite{bendikov_chem_rev_2004,anthony_chem_rev_2006}
In spite of a long tradition of research,\cite{acene-historic} the
field of acenes has experienced a resurgence of interest in recent
years because they are also perceived as the building blocks for organic
electronic materials such as graphene nanoribbons.\cite{bendikov_chem_rev_2004,Anthony2008}

Although pentacene, has excellent optical and transport properties,
however, it is conceivable that the longer acenes could have even
more attractive properties, with possible applications in the field
of nanotechnology.\cite{Anthony2008} As the size of the longer acenes
approaches the nanometer scale, their reactivity also increases, and,
therefore, it has been difficult to synthesize them from heptacene
onwards.\cite{Schmidt1980} Recently, many efforts have been made
to synthesize the longer acenes, \emph{e.g.}, heptacene and its functionalized
derivatives have been synthesized by several workers.\cite{Payne2005,mondal_neckers_JACS2006,bettinger_neckers_hept_charge_sep_Chem_comm_2007,Chun2008,mondal_bettinger_JACS2009}
By using cryogenic matrix-isolation technique, and a protection group
strategy, octacene and nonacene have been synthesized by Tönshoff
and Bettinger\cite{Tonshoff2010}, while Kaur \emph{et al}.\cite{Kaur_non_JACS2010}
prepared functionalized nonacene.

While all the known oligoacenes ranging from naphthalene to hexacene
have a singlet ground state, some years back Angliker \emph{et al}.\cite{Angliker}
predicted nonacene to be the first linear acene with the triplet state
($1^{3}B_{2u}$) as the ground state, instead of the singlet one ($1^{1}A_{g}$).
Their prediction was based upon: (a) an extrapolation of the available
experimental values of the singlet-triplet gap of the shorter acenes,
and (b) theoretical calculations of the triplet states ($1^{3}B_{2u}$
) of acenes using the using the singles configuration interaction
(SCI) method, and the Pariser-Parr-Pople (PPP) model Hamiltonian.\cite{Angliker}
This was an interesting prediction because, if true, it could open
the possibilities of magnetic applications of longer acenes. The singlet-triplet
crossover in long acenes, predicted in this early work of Angliker
\emph{et al}.\cite{Angliker}, was also verified by Houk \emph{et
al.}\cite{Houk} based upon first-principles density-functional-theory
(DFT) based calculations. However, subsequent theoretical investigations
have predicted long acenes to have singlet ground states. They include
PPP model based density-matrix renormalization group (DMRG) calculations
of Raghu \emph{et al}.,\cite{Ramasesha-acene-2002}, valence-bond
theory based work of Gao \emph{et al.,\cite{gao_st_gap_2002}} first-principles
DFT calculations of Bendikov \emph{et al}.,\cite{Bendikov}\emph{
ab initio }DMRG calculations of Hachmann \emph{et al}.,\cite{Hachmann2007}
DFT-based work of Jiang and Dai,\cite{jiang_dai_jpc} and first-principles
coupled-cluster calculations of Hajgato \emph{et al.}\cite{FPA2011}
Recently, based upon optical absorption experiments, Tönshoff and
Bettinger\cite{Tonshoff2010} demonstrated that nonacene has a singlet
ground state, thus contradicting the prediction of Angliker \emph{et
al}.\cite{Angliker} empirically. Motivated by this experimental finding,
we decided to perform a systematic theoretical investigation of the
electronic structure of the ground and low-lying excited states of
longer acenes, with an emphasis on the singlet-triplet gap, and their
optical properties. In order to realize the possible potential of
longer acenes in nanotechnology, a deep theoretical understanding
of their electronic structure is very important. For our calculations,
we adopt a methodology based upon the PPP model Hamiltonian, along
with large-scale multi-reference singles-doubles configuration interaction
(MRSDCI) approach.

First we benchmark our PPP-MRSDCI methodology by performing calculations
of the singlet-triplet gaps of acenes ranging from naphthalene to
decacene, and obtain results in very good quantitative agreement with
those obtained by other wave-function-based approaches.\cite{Ramasesha-acene-2002,Hachmann2007,FPA2011}
Next, with the aim of understanding the experiments of Tönshoff and
Bettinger,\cite{Tonshoff2010} we compute the optical absorption spectra
of octacene, nonacene, and decacene, both for the singlet and triplet
manifolds. We discover that the results of our singlet optical absorption
calculations are in excellent agreement with the experimental results
for octacene and nonacene,\cite{Tonshoff2010} leading us to conclude
that the ground state in nonacene is of singlet multiplicity, against
the predictions of Angliker \emph{et al}.\cite{Angliker} Although,
the experimental results for decacene do not exist as of now, however,
based upon the excellent quantitative agreement we obtain with the
measured absorption spectra of octacene and nonacene, we predict that
the ground state of decacene is also a singlet. Our computed optical
absorption spectra for the triplet manifolds of octacene, nonacene,
and decacene predict two major well-separated peaks polarized along
the long axis, which can be tested in future experiments on longer
acenes.

\section{{\normalsize Theory}}

The schematic structures of higher polyacenes studied in this work
are shown in Figure \ref{fig:acene}. The molecule is assumed to lie
in the $xy$-plane, with the conjugation direction taken to be along
the $x$-axis. The carbon-carbon bond length has been fixed at 1.4
Å, and all bond angles have been taken to be 120$^{o}$. It can be
noted that these structures can also be seen as two polyene chains
of suitable lengths, coupled together along the $y$-direction. The
reason of choosing this symmetric geometry, against various other
possibilities has already been discussed in our earlier paper.\cite{sony-acene-lo,sony-acene-pa}
\textcolor{black}{However, in the supporting information, we do consider
an alternate geometry with nonuniform bond lengths,\cite{Bendikov}
and demonstrate that it leads to small quantitative changes in the
optical absorption spectrum, compared to the results obtained using
the uniform bond length. Thus, we conclude that small differences
in the ground state geometry lead to insignificant changes in the
optical absorption spectra of oligo acenes. }

\begin{center}
\begin{figure}[H]
\begin{centering}
\includegraphics[scale=0.35]{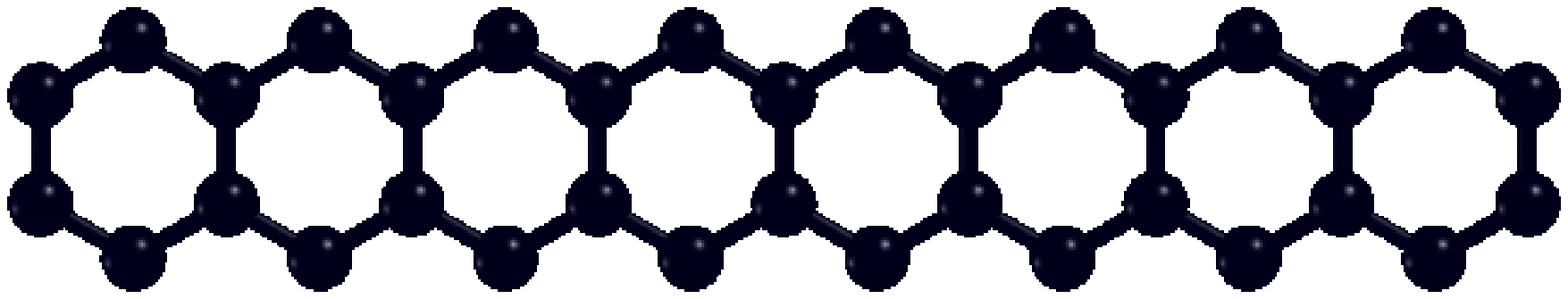} 
\par\end{centering}

\begin{centering}
(a) 
\par\end{centering}

\begin{centering}
\includegraphics[scale=0.4]{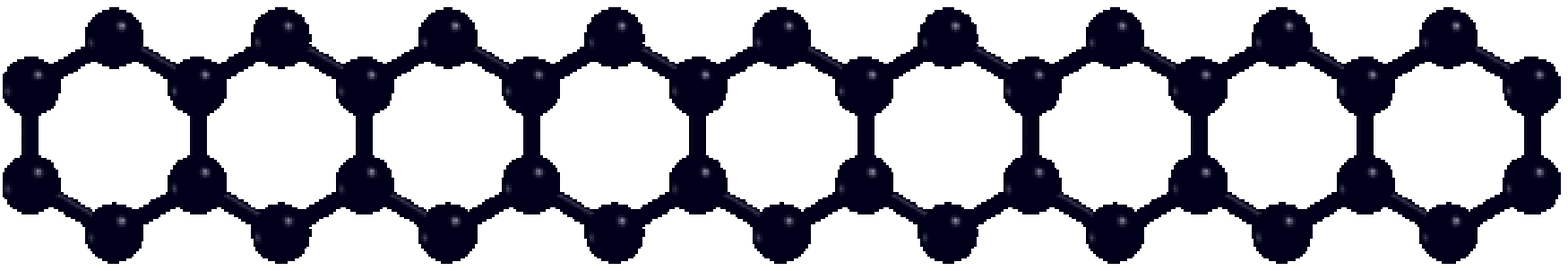} 
\par\end{centering}

\begin{centering}
(b) 
\par\end{centering}

\begin{centering}
\includegraphics[scale=0.45]{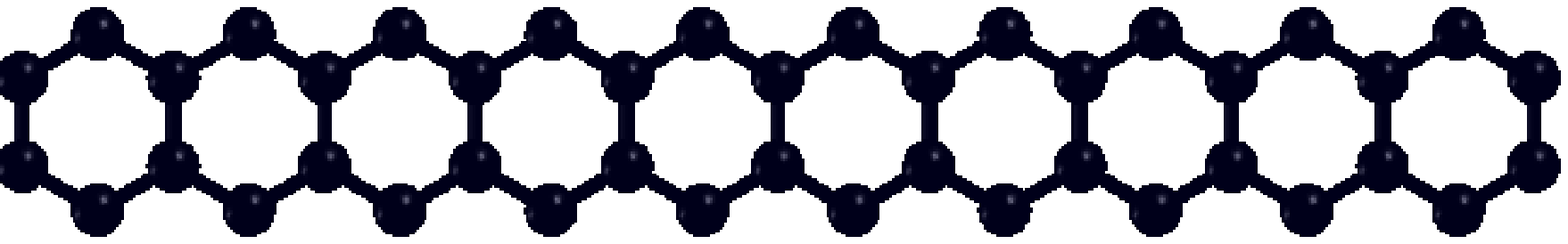} 
\par\end{centering}

\begin{centering}
(c) 
\par\end{centering}

\centering{}\caption{Schematic drawings of long acenes considered in this work, (a) octacene,
(b) nonacene and (c) decacene. The $x$ axis is assumed along the
conjugation direction, while the $y$ axis is perpendicular to it,
in the plane of the figure. \label{fig:acene}}
\end{figure}

\par\end{center}

The correlated calculations are performed using the PPP model Hamiltonian,
which can be written as

\[
H=H_{C_{1}}+H_{C_{2}}+H_{C_{1}C_{2}}+H_{ee},
\]
 where $H_{C_{1}}$ and $H_{C_{2}}$ are the one-electron Hamiltonians
for the carbon atoms located on the upper and the lower polyene like
chains, respectively. $H_{C_{1}C_{2}}$ is the one-electron hopping
between the two chains, and H$_{ee}$ depicts the electron-electron
repulsion. The individual terms can now be written as,

\[
H_{C_{1}}=-t_{0}\Sigma_{\langle k,k'\rangle}B_{k,k'},
\]
\[
H_{C_{2}}=-t_{0}\Sigma_{\langle\mathbf{\mu},\nu\rangle}B_{\mu,\nu},
\]

\[
H_{C_{1}C_{2}}=-t_{\perp}\Sigma_{\langle k,\mu\rangle}B_{k,\mu},
\]
 
\begin{eqnarray*}
H_{ee} & = & U\Sigma_{i}n_{i\uparrow}n_{i\downarrow}+\frac{1}{2}\Sigma_{i\neq j}V_{i,j}(n_{i}-1)(n_{j}-1).
\end{eqnarray*}
 In the equation above, $k$, $k'$ are carbon atoms on the upper
polyene chain, $\mu,\nu$ are carbon atoms located on the lower polyene
chain, while $i$ and $j$ represent all the atoms of the oligomer.
Symbol $\langle...\rangle$ implies nearest neighbors, and $B_{i,j}=\Sigma_{\sigma}(c_{i,\sigma}^{\dagger}c_{j,\sigma}+h.c.)$,
where $c_{i,\sigma}^{\dagger}\:\mbox{(}c_{i,\sigma}\mbox{)}$ denotes
the creation (annihilation) operator for a $\pi$ orbital of spin
$\sigma$, located on the $i$th carbon atom. Matrix elements $t_{0}$,
and $t_{\perp}$ depict one-electron hops. As far as the values of
the hopping matrix elements are concerned, we took $t_{0}=2.4$ eV
for all nearest neighbor hopping with $t_{\perp}=t_{0}$, consistent
with the undimerized ground state for polyacene argued by Raghu \emph{et
al}.\cite{Ramasesha-instability-12002}

The Coulomb interactions are parametrized according to the Ohno relationship,\cite{Ohno1964}
\[
V_{i,j}=U/\kappa_{i,j}(1+0.6117R_{i,j}^{2})^{1/2}\;\mbox{,}
\]

where, $\kappa_{i,j}$ depicts the dielectric constant of the system
which can simulate the effects of screening, $U$ is the on-site repulsion
term, and $R_{i,j}$ is the distance in \AA{}\ between the $i$th
and the $j$th carbon atoms. In the present work, we have performed
calculations using ``standard parameters''\cite{Ohno1964} with
$U=11.13$ eV and $\kappa_{i,j}=1.0$, as well as ``screened parameters''\cite{Chandross}
with $U=8.0$ eV and $\kappa_{i,j}=2.0$ ($i\neq j$) and $\kappa_{i,i}=1.0$.
The screened parameters employed here were devised by Chandross and
Mazumdar\cite{Chandross} with the aim of accounting for the inter
chain screening effects in phenylene based polymers. \textcolor{black}{However,
in the present case, we compare our results of singlet and triplet
absorption spectra of octacene and nonacene with the experimental
spectra measured by Bettinger and co-workers,\cite{Tonshoff2010}
with the oligomers located inside the argon matrix. Therefore, here
the purpose of screened parameters is to physically model the screening
due to the argon matrix.}

The starting point of the correlated calculations for the molecules
is the Restricted Hartree-Fock (RHF) calculations, employing the PPP
Hamiltonian, using a code developed in our group.\cite{priya_ppp}
All the resultant HF molecular orbitals are treated as active orbitals.
The single-reference CI calculations such as the full/quadruple configuration
interaction (FCI/QCI) were employed for shorter acenes, while the
multi-reference singles doubles configuration interaction (MRSDCI)
method was used for the longer ones. In particular, FCI method was
used for naphthalene and anthracene, the QCI method was employed for
the $^{1}A_{g}$ and $^{1}B_{2u}$ symmetries of tetracene and pentacene,
while for all other cases MRSDCI calculations were performed. The
MRSDCI method is a well-established approach for including electron-correlation
effects beyond the mean-field both for the ground and excited states
of molecular systems.\cite{peyerimhoff_CI,peyerimhoff_energy_CI}
We have used this approach extensively within the PPP model to study
the optical properties of a number of conjugated polymers,\cite{Shukla-PPV,Shukla-THG,Shukla-TPA,Shukla2}
and it can be briefly summarized as follows. After the RHF calculations
of the ground state $1^{1}A_{g}$, the CI calculation of the ground
state $1^{1}A_{g}$, and the excited states of symmetries $^{1}B_{2u}$
and $^{1}B_{3u}$ are performed by taking the lowest energy configuration
of the corresponding symmetry. They are  $H\rightarrow L$ for $B_{2u}$ and
 $H\rightarrow L+m$ and $H-m\rightarrow L$ ($m$ > 0 is an integer whose value 
depends on the length of the oligomer) for $B_{3u}$, where $H$ and $L$ 
corresponds to Highest Occupied Molecular Orbital
(HOMO) and Lowest Unoccupied Molecular Orbital (LUMO), respectively.
A similar approach is adopted for the triplet states as well, except
that one has to ensure that the spin multiplicity of the wave functions
is triplet. From the CI calculations, we obtain the eigenfunctions
and eigenvalues corresponding to the correlated ground and excited
states of the examined molecules. Using these eigenfunctions, the
dipole matrix elements between the ground state and various excited
states are computed. For the triplet states, the matrix elements are
calculated with respect to the lowest triplet state $1^{3}B_{2u}$.
These dipole matrix elements, along with the energies of the excited
states are, in turn, utilized to calculate linear (or triplet) optical
absorption spectrum. Important excited states contributing to various
peaks of the spectrum are identified, and a new set of MRSDCI calculations
are performed with an increased number of reference configurations
contributing both to the ground state, and the excited states, leading
to a new absorption spectrum. This procedure is iterated until the
computed spectrum converges satisfactorily.

\section{{\normalsize Results and Discussions}}

\label{sec:results}

In this section we present the results of our CI calculations on polyacenes
ranging from naphthalene to decacene examining their singlet-triplet
gaps, with the aim of determining the spin-multiplicity of their ground
state. Furthermore, we also present MRSDCI calculations on the optical
absorption of the long acenes, namely, octacene to decacene from their
lowest triplet states ($1^{3}B_{2u}$), and the lowest-singlet state
($1^{1}A_{g}$), and compare the results with the experimental ones,
where available.

\subsection{Singlet-Triplet gap}

We performed the first set of calculations to explore the singlet-triplet
($1^{1}A_{g}$-$1^{3}B_{2u}$) gap in oligoacenes using the singles-configuration-interaction
(SCI) method. The values of the singlet-triplet gap $\Delta E_{ST}=E(1^{3}B_{2u})-E(1^{1}A_{g})$
obtained from these calculations for oligoacenes up to acene-15 (acene-$n$
denotes an oligomer with $n$ benzene rings) are presented in Table
\ref{tab:RHF}. Examining the values of the singlet-triplet gap for
various oligomers, it is obvious that: (a) in the standard-parameter
based calculations, triplet state is never lower than the singlet
one, while, (b) with the screened parameters, the singlet and the
triplet states become nearly degenerate for acene-7, and the crossover
takes place starting with acene-8. In both sets of calculations, $\Delta E_{ST}$
first decreases with increasing $n$, and, subsequently begins to
increase, suggesting the inadequacy of the SCI method for longer acenes.
Nevertheless, our screened-parameter based SCI results appear to confirm
the essential prediction of Angliker \emph{et al}.\cite{Angliker}
that the singlet-triplet crossover will take place in oligoacenes
with the increasing lengths, although, they differ in details in that
Angliker \emph{et al.}\cite{Angliker} predicted the crossover from
acene-9 onwards. Although Angliker \emph{et al.}\cite{Angliker} also
used the PPP-SCI approach for their calculations, however, they used
a smaller value of the nearest-neighbor hopping matrix element $t=-2.318$
eV, and a Mataga-Nishimoto\cite{mataga_nishimoto} type Coulomb parameterization
$V_{ij}=1439.5/(132.8+R_{ij})$, which corresponds to an effective
value of $U=10.84$ eV. It is obvious that the results of our SCI-PPP
calculations differ from those of Angliker \emph{et al.}\cite{Angliker}
because of different values of parameters employed.  Furthermore,
another calculation, performed by Houk \emph{et al.},\cite{Houk}
also predicted the singlet-triplet crossover in oligoacenes. \textcolor{black}{However,
}\textcolor{black}{\emph{ab initio}}\textcolor{black}{{} DFT \cite{Houk}
can be unreliable for treating multi-reference correlation effects,
which are important in the excited states of conjugated systems.}
Based upon past calculations performed on other polymers such as \emph{trans}-polyacetylene,\cite{sumit-review}
it is a well-known fact that in order to be able to predict the correct
excited state orderings in conjugated polymers, it is very important
to account for the electron-correlation effects in an accurate manner.
Therefore, we decided to go beyond the SCI approach, and performed
large-scale CI calculations to explore the singlet-triplet ordering
in polyacenes. Before we present and discuss our results, we would
like to give a flavor to the reader as to the size of the CI calculations
performed. Table \ref{tab:no.ofconfig} presents the number of reference
states used in the MRSDCI calculations, and the total size of the
resultant CI matrix. \textcolor{black}{As mentioned in the previous
section, the number of references ($N_{ref}$) used in various MRSDCI
calculations was increased until acceptable convergence was achieved.
For example, the convergence of the excitation energies of $1^{1}B_{2u}^{+}$
and $1^{3}B_{2u}^{+}$ states of nonacene with respect to$N_{ref}$,
computed using the screened parameters, has been demonstrated explicitly
in the Supporting Information.}

The calculated values of the singlet-triplet gap are presented in
Table \ref{tab:sing_trip}, using both the standard and the screened
parameters, for the oligoacenes ranging from naphthalene to decacene.
The variation of the calculated values of the singlet-triplet gap,
and their comparison with the experimental and other theoretical calculations,
for the oligoacenes ranging from naphthalene to decacene, as a function
of the oligomer length, is presented in Figure \ref{fig:sing_trip}.
It is obvious both from the figure, that the excitation energy of
$1^{3}B_{2u}$ state decreases as the oligomer length increases. Nevertheless,
even for decacene the singlet-triplet gap is nonvanishing, and appears
to saturate as a function of the increasing chain length. As far as
comparison with the work of other authors is concerned, we observe
the followin\ref{fig:acene10_pa}g trends in Figure \ref{fig:sing_trip}:
(a) Our standard parameter results are in good quantitative agreement
with most other works, (b) in particular, our standard parameter results
are in excellent agreement with the PPP-DMRG results of Raghu \emph{et
al.},\cite{Ramasesha-acene-2002} and also in very good agreement
with the \emph{ab initio} DMRG results of Hachmann \emph{et al}.,\cite{Hachmann2007}
further vindicating our MRSDCI approach, and (c) our screened parameter
results predict smaller values of the singlet-triplet gap as compared
to other wave function based approaches for the shorter acenes, however,
for longer ones they are also in good agreement with other results.
However, DFT-based UB3LYP calculation performed by Bendikov \emph{et
al.}\cite{Bendikov} agree with other works for shorter acenes, but
for longer ones predict much smaller singlet-triplet gaps. We believe
that these smaller gaps obtained in the work of Bendikov \emph{et
al.},\cite{Bendikov} could be attributed to the well-known tendency
of DFT to underestimate the energy gaps. Therefore, based upon the
fact that for decacene our calculations predict a singlet-triplet
gap $>0.5$ eV, we conclude that even for longer acenes, the singlet
state $1^{1}A_{g}$ will be the ground state, and thus no singlet-triplet
crossover of the kind predicted by Angliker \emph{et al}.,\cite{Angliker}
occurs as per our calculations, and in agreement with all other works
except that of Houk \emph{et al}.\cite{Houk} Recently, Hajgato \emph{et
al.} \cite{FPA2011} performed first principles coupled cluster CCSD(T)
calculations on the singlet-triplet gaps of acenes ranging from octacene
to undecacene (acene-11), and their reported value of 0.58 eV for
octacene is in almost perfect agreement with our PPP-MRSDCI results
(\emph{cf}. Table \ref{tab:sing_trip}). However, for nonacene and
decacene their reported values 0.46 eV and 0.35 eV\cite{FPA2011},
respectively, are smaller than both our standard and screened parameter
values (\emph{cf}. Table \ref{tab:sing_trip}). Regarding the singlet-triplet
gap in the polyacene limit ($n\rightarrow\infty)$, DMRG-PPP work
of Raghu \emph{et al}.\cite{Ramasesha-acene-2002} predicted it to
be 0.53 eV, while Gao \emph{et al.,\cite{gao_st_gap_2002}} using
a spin Hamiltonian, estimated it to be 0.446 eV, both of which are
reasonably close to our standard/screened parameters values of 0.57/0.54
eV, computed for decacene.

Bendikov \emph{et al}.\cite{Bendikov} noted that the restricted singlet
density functional ground state of higher acenes would become unstable
due to its open-shell nature, with two unpaired electrons (a singlet
diradical\cite{Salem1972}) for acenes longer than hexacene. Based
upon \emph{ab initio} DMRG calculations on acenes in the range $n=2-12$,
Hachmann \emph{et al}.\cite{Hachmann2007} concluded that the ground
state wave functions for longer acenes were of the type of polyradical
singlets. In another DFT work, Jiang and Dai\cite{jiang_dai_jpc}
predicted the ground state of octacene and higher acenes to be antiferromagnetic
(in other words, open-shell singlet), but not necessarily a diradical.
Unlike the DFT calculations, in our approach, the many-body wave functions
of the ground state and the excited states of the studied oligomers
are available. Therefore, we decided to probe the nature of the ground
state of longer acenes to ascertain whether, or not, they exhibit
a polyradical character. The character of the many-body wave functions
of the $1^{1}A_{g}$ ground state, obtained in our best CI calculations
on oligomers ranging from naphthalene to decacene are presented in
Table \ref{tab:Open-shell}, for both the standard, and the screened
parameters. From the results it is obvious that the singlet ground
state of longer acenes begins to exhibit significant configuration
mixing. \textcolor{black}{With the increasing oligomer lengths, the
contributions of several doubly-excited configurations to the ground
state wave function increase.} Thus, based upon these results, we
conclude that the longer acenes studied in this work exhibit a tendency
towards a singlet ground state, with a significant diradical open-shell
character.

\begin{center}
\begin{table}[H]
\begin{centering}
\begin{tabular}{lcc}
\hline 
$n$  & \multicolumn{2}{c}{$\Delta E_{ST}$ (eV)}\tabularnewline
\hline 
 & std & scr\tabularnewline
\hline 
$2$ & $2.66$  & $2.07$\tabularnewline
$3$  & $1.63$  & $1.16$\tabularnewline
$4$  & $1.04$  & $0.63$\tabularnewline
$5$  & $0.68$  & $0.32$\tabularnewline
$6$  & $0.42$  & $0.12$\tabularnewline
$7$  & $0.34$  & $0.00$\tabularnewline
$8$  & $0.26$  & $-0.06$\tabularnewline
$9$  & $0.23$  & $-0.10$\tabularnewline
$10$  & $0.24$  & $-0.11$\tabularnewline
$11$  & $0.26$  & $-0.12$\tabularnewline
$15$  & $0.41$  & $-0.06$\tabularnewline
\hline 
\end{tabular}
\par\end{centering}

\raggedright{}\caption{The singlet-triplet gap ($\Delta E_{ST}=E(1^{3}B_{2u})-E(1^{1}A_{g})$)
of acene-$n$, computed using the SCI method, and the standard (std),
and the screened (scr) Coulomb parameters.\label{tab:RHF}}
\end{table}

\par\end{center}

\begin{center}
\begin{table}[H]
\caption{The number of reference configurations ($N_{ref})$ and the total
number of configurations $(N_{total})$ involved in the MRSDCI (or
FCI or QCI, where indicated) calculations, for different symmetry
subspaces of various oligoacenes. \label{tab:no.ofconfig}}

\vspace{0.25cm}

\begin{centering}
\begin{tabular}{ccccccc}
\hline 
$n$  & \multicolumn{2}{c}{$^{1}A_{g}$} & \multicolumn{2}{c}{\textsuperscript{1}$B_{2u}$} & \multicolumn{2}{c}{\textsuperscript{3}$B_{2u}$}\tabularnewline
\hline 
 & $N_{ref}$  & $N_{total}$  & $N_{ref}$  & $N_{total}$  & $N_{ref}$  & $N_{total}$\tabularnewline
\hline 
$2$  & $1^{a}$  & $4936^{a}$  & $1^{a}$  & $4794^{a}$  & $1^{a}$  & $4816{}^{a}$\tabularnewline
$3$  & $1^{a}$  & $623576{}^{a}$  & $1^{a}$  & $618478{}^{a}$  & $1^{a}$  & $620928{}^{a}$\tabularnewline
$4$  & $1^{b}$  & $193538{}^{b}$  & $1^{b}$  & $335325{}^{b}$  & $100^{c}$  & $323063^{c}$\tabularnewline
 &  &  &  &  & $86^{d}$  & $319005^{d}$\tabularnewline
$5$  & $1^{b}$  & $1002597{}^{b}$  & $1^{b}$  & $1707243{}^{b}$  & $52^{c,d}$  & $581702^{c,d}$\tabularnewline
$6$  & $100^{c}$  & $1110147{}^{c}$  & $100^{c}$  & $1173212{}^{c}$  & $65^{c}$  & $1461526^{c}$\tabularnewline
 & $100^{d}$  & $1177189{}^{d}$  & $100^{d}$  & $1328252{}^{d}$  & $63^{d}$  & $1551590^{d}$\tabularnewline
$7$  & $35{}^{c}$  & $856788{}^{c}$  & $30{}^{c}$  & $674925^{c}$  & $33{}^{c}$  & $1369624{}^{c}$\tabularnewline
 & $22{}^{d}$  & $615590{}^{d}$  & $30{}^{d}$  & $850627{}^{d}$  & $29{}^{d}$  & $1300948{}^{d}$\tabularnewline
$8$  & $18{}^{c}$  & $768641{}^{c}$  & $14^{c}$  & $509119{}^{c}$  & $19^{c}$  & $1066355^{c}$\tabularnewline
 & $12{}^{d}$  & $540651{}^{d}$  & $4{}^{d}$  & $145978{}^{d}$  & $14{}^{d}$  & $918645^{d}$\tabularnewline
$9$  & $13^{c}$  & $959737{}^{c}$  & $13{}^{c}$  & $769387{}^{c}$  & $18^{c}$  & $1626229^{c}$\tabularnewline
 & $12^{d}$  & $871397^{d}$  & $3{}^{d}$  & $186651{}^{d}$  & $12^{d}$  & $1152071{}^{d}$\tabularnewline
$10$  & $11^{c}$  & $1202681^{c}$  & $12{}^{c}$  & $1192394^{c}$  & $15{}^{c}$  & $1735352{}^{c}$\tabularnewline
 & $11{}^{d}$  & $1199887{}^{d}$  & $3{}^{d}$  & $270187{}^{d}$  & $10{}^{d}$  & $1318156{}^{d}$\tabularnewline
\hline 
\end{tabular}
\par\end{centering}

$^{a}$FCI method with standard as well as screened parameters,

$^{b}$QCI method with standard as well as screened parameters,

$^{c}$using standard parameters,

$^{d}$using screened parameters. 
\end{table}

\par\end{center}

\begin{center}
\begin{table}[H]
\caption{For various oligomers, the singlet-triplet gaps ($\Delta E_{ST}=E(1^{3}B_{2u})-E(1^{1}A_{g})$)
\label{tab:sing_trip} obtained from large-scale CI calculations (\emph{cf}.
Table \ref{tab:no.ofconfig}), employing the PPP model Hamiltonian,
and the standard (std) and screened (scr) parameters. }

\centering{}%
\begin{tabular}{ccc}
\hline 
$n$  & $\Delta E_{ST}$ (eV) & \tabularnewline
\hline 
 & std  & scr \tabularnewline
$2$  & $2.53$  & $2.11$ \tabularnewline
$3$  & $1.73$  & $1.48$ \tabularnewline
$4$  & $1.25$  & $1.11$ \tabularnewline
$5$  & $0.99$  & $0.93$ \tabularnewline
$6$  & $0.87$  & $0.85$ \tabularnewline
$7$  & $0.73$  & $0.69$ \tabularnewline
$8$  & $0.68$  & $0.58$ \tabularnewline
$9$  & $0.60$  & $0.56$ \tabularnewline
$10$  & $0.57$  & $0.54$ \tabularnewline
\hline 
\end{tabular}
\end{table}

\par\end{center}

\begin{center}
\begin{figure}[H]
\centering{}\includegraphics[width=8cm]{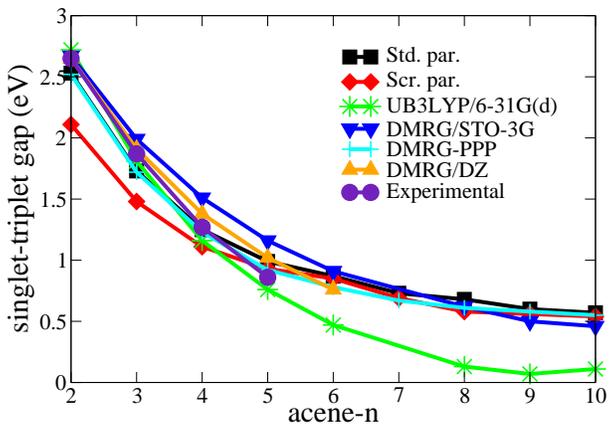}\caption{(Color online) Singlet-triplet energy gap as a function of the acene
lengths. \label{fig:sing_trip}}
\end{figure}

\par\end{center}

\begin{center}
\begin{table}[H]
\begin{centering}
\caption{Magnitudes of the coefficients of the closed shell ($CS$) doubly
excited virtual configuration $\vert H\rightarrow L;\, H\rightarrow L\rangle$,
and the open shell ($OS$) doubly excited virtual configuration $\vert H\rightarrow L;\, H-1\rightarrow L+1\rangle$
to the singlet $1^{1}A_{g}$ ground state CI wave functions of acene-$n$,
obtained using the standard (std), and the screened (scr) parameters
in the PPP model. In the preceding expression $H$ stands for the
highest occupied molecular orbital (HOMO), while $L$ stands for lowest
unoccupied molecular orbital (LUMO). \label{tab:Open-shell}}

\par\end{centering}

\centering{}\centering{}%
\begin{tabular}{ccccc}
\hline 
\multicolumn{1}{c}{$n$} & \multicolumn{2}{c}{\textcolor{black}{$CS$}} & \multicolumn{2}{c}{\textcolor{black}{$OS$}}\tabularnewline
\hline 
 & \textcolor{black}{std } & \textcolor{black}{scr } & \textcolor{black}{std } & \textcolor{black}{scr}\tabularnewline
\hline 
$2$  & \textcolor{black}{$0.114$ } & \textcolor{black}{$0.143$ } & \textcolor{black}{$0.145$ } & \textcolor{black}{$0.126$}\tabularnewline
$3$  & \textcolor{black}{$0.148$ } & \textcolor{black}{$0.168$ } & \textcolor{black}{$0.132$ } & \multicolumn{1}{c}{\textcolor{black}{$0.109$}}\tabularnewline
$4$  & \textcolor{black}{$0.173$ } & \textcolor{black}{$0.179$ } & \textcolor{black}{$0.119$ } & \multicolumn{1}{c}{\textcolor{black}{$0.134$}}\tabularnewline
$5$  & \textcolor{black}{$0.191$ } & \textcolor{black}{$0.186$ } & \textcolor{black}{$0.134$ } & \multicolumn{1}{c}{\textcolor{black}{$0.140$}}\tabularnewline
$6$  & \textcolor{black}{$0.198$ } & \textcolor{black}{$0.184$ } & \textcolor{black}{$0.144$ } & \multicolumn{1}{c}{\textcolor{black}{$0.142$}}\tabularnewline
$7$  & \textcolor{black}{$0.234$ } & \textcolor{black}{$0.269$ } & \textcolor{black}{$0.168$ } & \multicolumn{1}{c}{\textcolor{black}{$0.188$}}\tabularnewline
$8$  & \textcolor{black}{$0.246$ } & \textcolor{black}{$0.289$ } & \textcolor{black}{$0.186$ } & \multicolumn{1}{c}{\textcolor{black}{$0.202$}}\tabularnewline
$9$  & \textcolor{black}{$0.263$ } & \textcolor{black}{$0.294$ } & \textcolor{black}{$0.194$ } & \multicolumn{1}{c}{\textcolor{black}{$0.211$}}\tabularnewline
$10$  & \textcolor{black}{$0.266$ } & \textcolor{black}{$0.297$ } & \textcolor{black}{$0.203$ } & \multicolumn{1}{c}{\textcolor{black}{$0.218$}}\tabularnewline
\hline 
\end{tabular}
\end{table}

\par\end{center}

\subsection{Singlet Linear Optical Absorption calculations}

In an earlier work in our group, we had reported the calculations
of linear optical absorption spectra of oligoacenes ranging from naphthalene
to heptacene.\cite{sony-acene-lo} In this work, we extend our calculations
to longer acenes, and present the calculations of linear optical absorption
in octacene, nonacene and decacene. In Figures \ref{fig:acene8_lo},
\ref{fig:acene9_lo} and \ref{fig:acene10_lo}, we present the singlet
linear absorption spectra of these oligoacenes from their $1^{1}A_{g}^{-}$
singlet state computed using the standard parameters and the screened
parameters. As per dipole selection rules for the $D_{2h}$ symmetry,
allowed one-photon transitions from the $1^{1}A_{g}^{-}$ state occur
to $^{1}B_{2u}^{+}\:(^{1}B_{3u}^{+})$ type states via short-axis
(long-axis) polarized photons, where we assume that the short (long)
axis correspond to $y$ ($x$) directions. The essential states contributing
to the linear absorption spectra of various acenes are depicted in
Figure \ref{fig:lo}. The many-particle wave functions of the excited
states contributing to various peaks in the spectra are presented
in \textcolor{black}{Tables \ref{tab:acene-8}-\ref{tab:acene-10-scr}
of the Supporting Information.} While plotting the absorption spectra,
we have restricted ourselves to states which lie below 6 eV, the estimated
value of the ionization potentials of the long acenes.\cite{ionization_acenes}

\begin{center}
\begin{figure}[H]
 \centering{}\includegraphics[width=6cm]{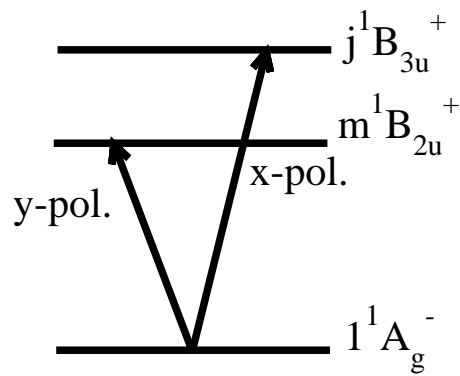}\caption{Diagram of the essential states involved in the singlet linear optical
absorption in oligoacenes and their polarization characteristics.
The arrows connecting two states imply optical absorption, with polarization
directions stated next to them. Location of states is not up to scale.
\label{fig:lo}}
\end{figure}

\par\end{center}

A peak-by-peak detailed description of the computed singlet absorption
spectra is provided in the Supporting Information. Here we list the
salient features of the calculated linear optical absorption of octacene,
nonacene, and decacene: 
\begin{enumerate}
\item Quantitatively speaking, screened parameter spectra are redshifted
as compared to the standard parameter ones. 
\item Most of the intensity is concentrated in the $x$-polarized (long-axis
polarized) spectra originating from the absorption into the $^{1}B_{3u}^{+}$
type of states, while the $y$-polarized (short-axis polarized) absorption
into the $^{1}B_{2u}^{+}$ type states is comparatively weak. However,
a closer examination reveals that most of the intensity in the $x$-polarized
spectrum is derived from the single transition to an $n^{1}B_{3u}^{+}$
state (or states which split away from it). If we ignore this transition
then the short- and long-axis polarized spectra are of comparable
intensity. This aspect of the singlet linear absorption spectrum of
long acenes is consistent with what is also observed in the shorter
acenes.\cite{sony-acene-lo} 
\item The first peak corresponds to the $y$-polarized transition, to the
$1{}^{1}B_{2u}^{+}$ excited state of the system. The most important
configuration contributing to the many-particle wave function of the
state corresponds to $\vert H\rightarrow L\rangle$ excitation, irrespective
of the choice of the Coulomb parameters employed in the PPP model. 
\item We have observed that the most intense absorption for the oligoacenes
is through an $x$-polarized photon to a $^{1}B_{3u}$ state, irrespective
of the Coulomb parameters employed in the calculations. For acene-$n$,
the many-particle wave function of this state exhibits the following
general features: (a) for the standard parameter case, single excitations
$\vert H\rightarrow L+n/2-1\rangle+c.c.$, for $n\equiv\mbox{even}$,
and $\vert H\rightarrow L+(n-1)/2\rangle+c.c.$, for $n\equiv\mbox{odd}$,
dominate the wave function, while (b) with screened parameters the
dominant configurations are single excitations $\vert H\rightarrow L+n/2\rangle+c.c.$
for $n\equiv\mbox{even}$, and $\vert H\rightarrow L+(n-1)/2\rangle+c.c.$
for $n\equiv\mbox{odd}$. The aforesaid difference between the standard
and the screened parameters is because of different energetic ordering
of the symmetries of the molecular orbitals for the standard and screened
parameters. 
\end{enumerate}
\begin{center}
\begin{figure}[H]
\caption{Singlet linear absorption spectra of octacene computed using: (a)
standard parameters and (b) screened parameters. A uniform line width
of 0.1 eV was assumed while plotting the spectra. The subscripts attached
to the peak labels indicate the polarization directions $x$ and $y$.
\label{fig:acene8_lo}}

\vspace{0.5cm}

\centering{}\includegraphics[width=8cm]{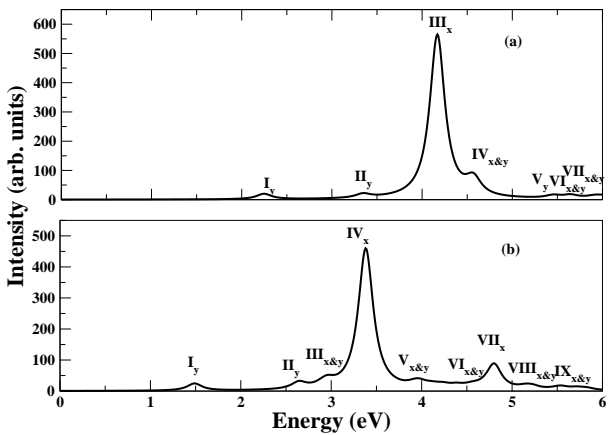} 
\end{figure}

\par\end{center}

\begin{center}
\begin{figure}[H]
\caption{Singlet linear absorption spectra of nonacene computed using: (a)
standard parameters and (b) screened parameters. The rest of the information
is same as in the caption of Figure \ref{fig:acene8_lo}.\label{fig:acene9_lo}}

\vspace{0.5cm}

\centering{}\includegraphics[width=8cm]{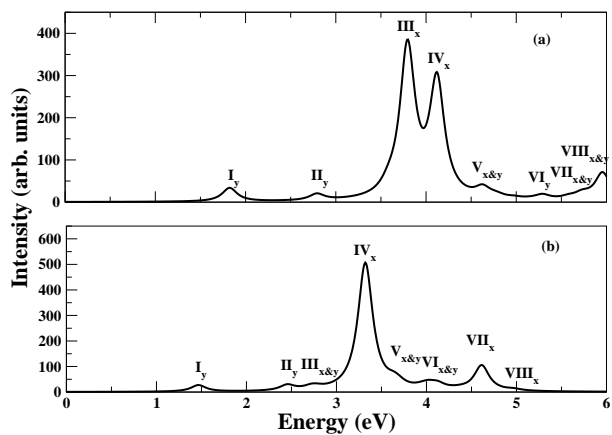} 
\end{figure}

\par\end{center}

\begin{center}
\begin{figure}[H]
\caption{Singlet linear absorption spectra of decacene computed using: (a)
standard parameters and (b) screened parameters. The rest of the information
is same as in the caption of Figure \ref{fig:acene8_lo}. \label{fig:acene10_lo}}

\vspace{0.5cm}

\centering{}\includegraphics[width=8cm]{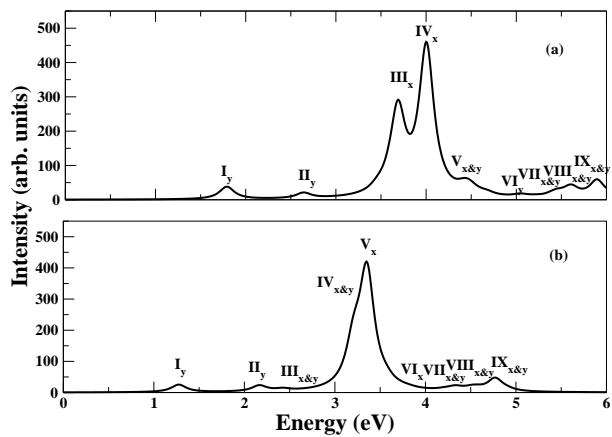} 
\end{figure}

\par\end{center}

\subsection{Triplet optical absorption calculations }

\begin{center}
\begin{figure}[H]
\begin{centering}
\includegraphics[clip,width=8cm]{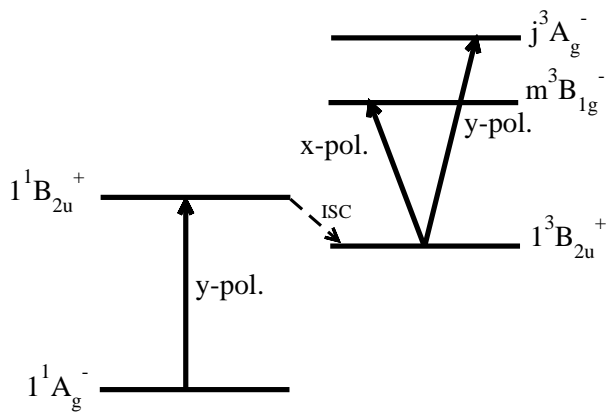} 
\par\end{centering}

\caption{Diagram of the essential states involved in the triplet absorption
spectra of oligoacenes, and their polarization characteristics. The
arrows connecting two states imply optical absorption, with polarization
directions stated next to them. Location of states is not up to scale.
ISC refers to inter-system crossing.\label{fig:PA}}
\end{figure}

\par\end{center}

In optical absorption experiments, one can probe the triplet excited
states because frequently the first singlet excited state $S_{1}$
($1^{1}B_{2u}^{+}$, in the present case) decays to the first triplet
excited state $T_{1}$ ($1^{3}B_{2u}^{+}$, in the present case) located
below $S_{1}$, through nonradiative inter-system crossing (ISC\emph{),
}as shown in Figure \ref{fig:PA}. Once the system reaches the triplet
manifold, normal optical absorption experiments can be performed to
probe higher triplet states. In the present work, we restrict ourselves
to the triplet one-photon absorption spectra of octacene, nonacene,
and decacene from their $1^{3}B_{2u}^{+}$ state, computed using the
MRSDCI method. For the case of oligoacenes, as per electric-dipole
selection rules of the $D_{2h}$ point group, the long-axis ($x$-axis)
polarized photons cause transitions from the $1^{3}B_{2u}^{+}$ to
$^{3}B_{1g}^{-}$ type of states, while the short-axis ($y$-axis)
polarized ones lead to the $^{3}A_{g}^{-}$ type states (\emph{cf}.
Figure \ref{fig:PA}). The calculated triplet absorption spectra of
these oligoacenes are displayed in Figures \ref{fig:acene8_pa}, \ref{fig:acene9_pa}
and \ref{fig:acene10_pa}, while the wave functions of the excited
states contributing to various peaks in the spectra are presented
in \textcolor{black}{Tables \ref{tab:acene-8-pa}-\ref{tab:acene-10-pa-scr}
of the Supporting Information.} While plotting the triplet absorption
spectra, we have been careful to include only those states which lie
below 6 eV excitation energy (with respect to the $1^{1}A_{g}^{-}$
state), which is the estimated value of the ionization potentials
of the long acenes.\cite{ionization_acenes} A detailed description
of the characteristics of various peaks in the calculated triplet
absorption spectra of octacene, nonacene, and decacene, is presented
in the Supporting Information. Below we discuss the salient features
of our results: 
\begin{enumerate}
\item Similar to the case of singlet absorption, screened parameter spectra
are red shifted as compared to the standard parameter ones. 
\item Most of the intensity is concentrated in the $x$-polarized (long-axis
polarized) spectra corresponding to the absorption into the $^{3}B_{1g}^{-}$
type of states, while the $y$-polarized absorption into the $^{3}A_{g}^{-}$
type states is very faint. 
\item From Figures \ref{fig:acene8_pa}, \ref{fig:acene9_pa}, and \ref{fig:acene10_pa}
it is obvious that the triplet absorption spectrum is dominated by
two intense $x$-polarized peaks which are well separated in energy
($>$2 eV), irrespective of the oligoacene in question, or the Coulomb
parameters employed in the calculations. The first of these peaks
is peak I in all the cases, while the second one is either peak IV
or V, depending upon the oligoacene, or the Coulomb parameters employed.
In the standard parameter based calculations, peak I is always the
second most intense peak, while the second of these peaks (IV or V)
is the most intense. In the screened parameter calculations the situation
is exactly the reverse, with peak I being the most intense, while
peak IV or V being the second most intense peak of the spectrum. Peak
I always corresponds to the $1^{3}B_{1g}^{-}$ excited state of the
system, whose wave function is dominated by the single excitations
$\vert H\rightarrow L+1\rangle+c.c.$, irrespective of the oligoacene
in question, or the choice of Coulomb parameters. The second intense
peak (IV or V) corresponds to a higher $^{3}B_{1g}^{-}$ type state
of acene-$n$, whose wave function mainly consists of: (a) for the
standard parameters, double excitations $\vert H\rightarrow L;\, H-(n/2-1)\rightarrow L\rangle+c.c.$
for $n\equiv\mbox{even}$, and $\vert H\rightarrow L;\, H-(n-1)/2\rightarrow L\rangle+c.c.$
for the $n\equiv\mbox{odd}$ case, while (b) for the screened parameters,
double excitations $\vert H\rightarrow L;\, H-n/2\rightarrow L\rangle+c.c.$
for $n\equiv\mbox{even}$, and $\vert H\rightarrow L;\, H-(n-1)/2\rightarrow L\rangle+c.c.$
for the $n\equiv\mbox{odd}$ case. This difference between the standard
and the screened parameter results is because of the different energetic
ordering of the molecular orbital symmetries in longer acenes, for
the two sets of Coulomb parameters. 
\item Other peaks in the spectrum correspond to either $x$ or $y$ polarized
transitions to the higher excited states of the system, which are
described in detail in the Supporting Information.
\end{enumerate}
\begin{center}
\begin{figure}[H]
\caption{Triplet absorption spectra of octacene from the $1^{3}B_{2u}^{+}$
state computed using: (a) standard parameters and, (b) screened parameters.
A uniform line width of 0.1 eV was assumed while plotting the spectra.
The subscripts attached to the peak labels indicate the polarization
directions $x$ and $y$. \label{fig:acene8_pa}}

\vspace{0.5cm}

\centering{}\includegraphics[width=8cm]{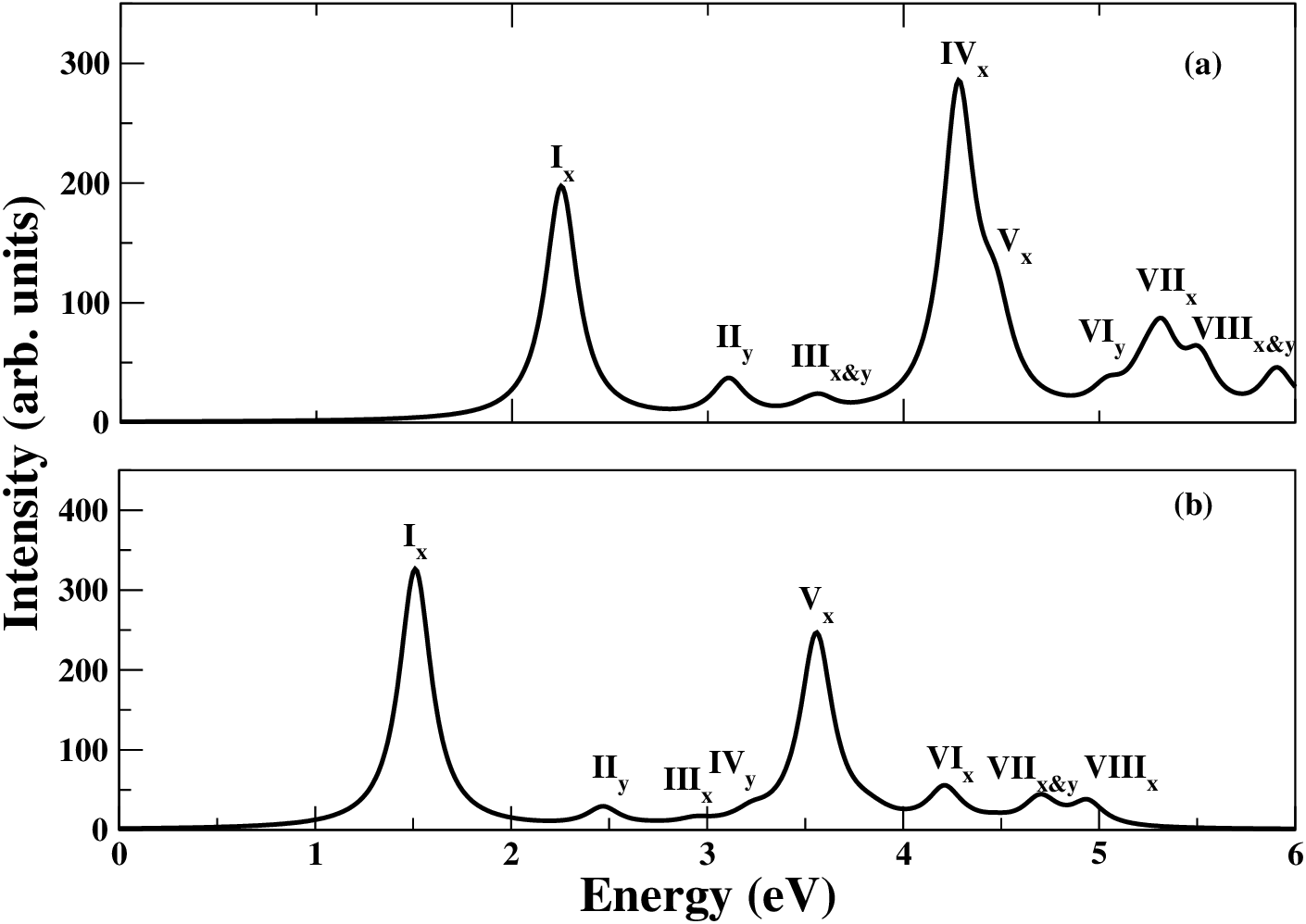} 
\end{figure}

\par\end{center}

\begin{center}
\begin{figure}[H]
\caption{Triplet absorption spectra of nonacene from the $1^{3}B_{2u}^{+}$
state computed using: (a) standard parameters and, (b) screened parameters.
The rest of the information is same as in the caption of Figure \ref{fig:acene8_pa}.\label{fig:acene9_pa}}

\vspace{0.5cm}

\centering{}\includegraphics[width=8cm]{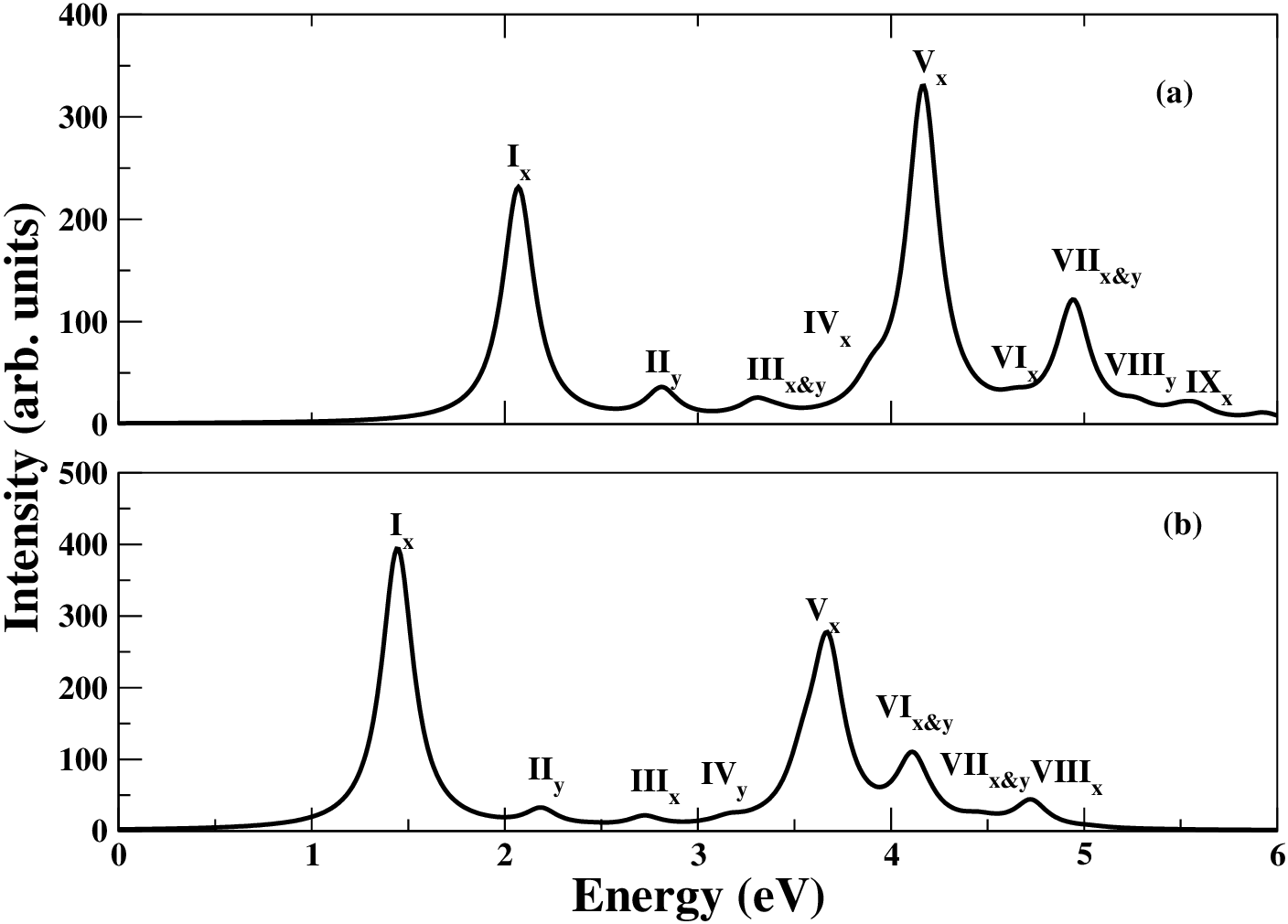} 
\end{figure}

\par\end{center}

\begin{center}
\begin{figure}[H]
\caption{Triplet absorption spectra of decacene from the $1^{3}B_{2u}^{+}$
state computed using: (a) the standard parameters and, (b) the screened
parameters. The rest of the information is same as in the caption
of Figure \ref{fig:acene8_pa}. \label{fig:acene10_pa}}

\vspace{0.5cm}

\centering{}\includegraphics[width=8cm]{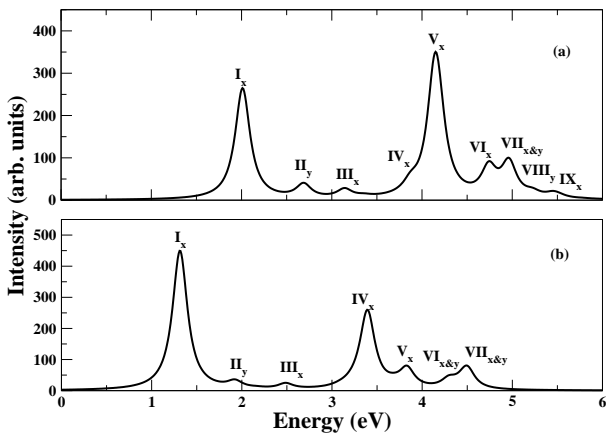} 
\end{figure}

\par\end{center}

\subsection{Comparison of singlet and triplet absorption calculations of octacene
and nonacene with the experimental results}

In Table \ref{tab:oct-non-expt}, the results of the experimental
absorption of octacene and nonacene, reported by Tönshoff and Bettinger,\cite{Tonshoff2010}
have been presented, showing the excitation energies, relative intensities,
and tentative assignments of the UV/vis electronic absorptions. Next,
we compare our computed singlet and triplet absorption results, with
the expermental ones, in terms of the peak energies (in eV) and relative
intensities/oscillator strength. The relative oscillator strength
(ROS) in the computed absorption spectra is the ratio of the calculated
oscillator strength of a given state, with respect to that of the
most intense state. The relative oscillator strengths of the peaks
in the calculated absorption spectra are compared with the reported
relative intensities (RI) of the corresponding peaks in the experimental
spectra.

\subsubsection{Octacene}

On comparing the experimental values of the energies of the absorption
peak of octacene from the Table \ref{tab:oct-non-expt}a, with our
calculated values for the singlet case (\textcolor{black}{see Tables
\ref{tab:acene-8}-\ref{tab:acene-8-scr}, Supporting Information}),
the lowest state $S_{1}$(1.54 - 1.86 eV, RI = 0.025 for the 1.54
eV) of the experimental results matches nicely with the first peak
at 1.49 eV (ROS = 0.050) obtained in the screened parameter calculations.
We note that the first peak of our standard parameter calculations
is located at 2.24 eV, and is thus significantly higher than the experimental
value. The second state of the experimental results $S_{2}$ (2.40
eV, RI = 0.02) again agrees much better with the screened parameter
result, 2.65 eV (ROS = 0.047) than with the standard parameter one
located at 3.34 eV. The third state of the experimental results, $S_{3}$
(2.54 eV, RI = 0.050) has a better agreement with the third state
of the screened parameter spectrum 2.97 eV (ROS = 0.046), than with
corresponding standard parameter peak at 4.17 eV. The most intense
peak in the experimental results corresponds to $S_{7}$ (3.68 - 3.78
eV, RI = 1.0) lies in between the highest peak of the screened and
standard parameter calculations located at 3.38 eV (ROS = 1.0), and
4.17 eV (ROS = 1.0), respectively. We note that with the increasing
peak energy, the agreement between the experiments and theory begins
to deteriorate. Nevertheless, even in the worst case scenario of the
most intense peak ($S_{7}$ of the experiments), the disagreement
between the screened parameter results and the experiments is below
10\%, as far as the peak location is concerned. This disagreement,
which is fairly acceptable from a quantitative aspect, is possibly
due to the reduced computational accuracy of our approach because
of the high energies of the states involved. As far as the comparison
of the experimental results with the triplet absorption calculations
(\textcolor{black}{see Tables \ref{tab:acene-8-pa}-\ref{tab:acene-8-pa-scr},
Supporting Information)} is concerned, the first peaks in the computed
triplet spectra, 3.04 eV (ROS = 0.839) for the standard parameter,
and 1.97 eV (ROS = 1.0) for the screened parameters, disagree completely
with the location of $S_{1}$, as well as its relative intensity in
the experimental spectrum. The same trend holds for the higher excited
states as well. Thus, we conclude that the experimental absorption
spectrum of octacene is indeed from the $1^{1}A_{g}^{-}$ ground state
of the system, confirming yet again that the ground state of octacene
has the singlet multiplicity.

We also note that Raghu \emph{et al}.\cite{Ramasesha-acene-2002}
predicted a very large optical gap of 2.60 eV for octacene, based
upon their DMRG calculations, employing the PPP Hamiltonian.

\subsubsection{Nonacene}

Similarly, comparing the experimental values of the energies of the
absorption peak of nonacene from the Table \ref{tab:oct-non-expt}b,
with our calculated values for the singlet case \textcolor{black}{(see
Tables \ref{tab:acene-9}-\ref{tab:acene-9-scr}, Supporting Information}),
the lowest state $S_{1}$ (1.43 - 1.62 eV, RI = 0.020 for 1.43 eV)
of the experimental results matches nicely with the first peak of
1.46 eV (ROS = 0.051) obtained in the screened parameter calculations,
than the corresponding one for the standard parameter case located
at 1.82 eV. The second state of the experimental results, $S_{3}$
(2.33 eV, RI = 0.033) again agrees much better with the second peak
of the screened parameter result, 2.45 eV (ROS = 0.043) than with
the standard parameter one located at 2.79 eV. The third state of
the experimental results, $S_{4}$ (2.50 eV, RI = 0.023) has a better
agreement with the third state of the screened parameter spectrum
2.71 eV (ROS = 0.020) than with the corresponding standard parameter
peak at 3.80 eV. The most intense peak in the experimental results
 $S_{9}$ (3.66 eV, RI = 1.0) lies in between the highest peak of
the screened and standard parameter calculations located at 3.32 eV
(ROS = 1.0), and 3.80 eV (ROS = 1.0), respectively. For this most
intense peak, the standard parameter results appear to agree slightly
better with the experiments, as compared to the screened ones. But,
keeping in mind the lower energy peaks discussed above, overall the
screened parameter based results have a much better agreement with
the experiments, just as in case of octacene. Similar to octacene,
we again note that with the increasing peak energy, the agreement
between the experiments and theory begins to deteriorate, which we
again attribute to the reduced computational accuracy for higher energies.
As far as the comparison of the experimental results with the triplet
absorption calculations (\textcolor{black}{see Tables \ref{tab:acene-9-pa}-\ref{tab:acene-9-pa-scr},
Supporting Information}) is concerned, the first peaks in the computed
triplet spectra, 2.56 eV (ROS = 0.785) for the standard parameter,
and 1.86 eV (ROS = 1.0) for the screened parameters, disagree completely
with the location of $S_{1}$, as well as its relative intensity in
the experimental spectrum. The same trend holds for the higher excited
states as well. Thus, we conclude that the experimental absorption
spectrum of nonacene is indeed from the $1^{1}A_{g}^{-}$ ground state
of the system, confirming yet again that the ground state of nonacene
has the singlet multiplicity. Again, we note that the DMRG based calculations
employing the PPP model, performed by Raghu \emph{et al}.\cite{Ramasesha-acene-2002}
predict an unrealistically large optical gap of 2.59 eV for nonacene.
Our values of the calculated optical gaps of octacene, nonacene, and
decacene are in very good agreement with the estimated optical gap
of polyacene $1.18\pm0.06$ eV, reported by Tönshoff and Bettinger.\cite{Tonshoff2010}

To conclude, the experimentalists are confident that the absorptions
which they observed are all from the ground states of octacene and
nonacene. Because, our singlet absorption spectra (as against the
triplet absorption) computed using the screened parameters are in
excellent agreement with those experiments, we conclude that the ground
states of octacene and nonacene are of singlet multiplicity ($1^{1}A_{g}$),
in perfect agreement with our singlet-triplet crossover calculations.
Although the experiments have not been performed on acenes longer
than nonacene, the trend emerging from our calculations leads us to
conclude that the ground state will be of singlet type in those systems
as well.

\begin{center}
\begin{table}[H]
\caption{The experimental values of the energies, relative intensities and
tentative assignments of the UV/vis electronic absorptions of (a)
octacene and (b) nonacene. \cite{Tonshoff2010} \label{tab:oct-non-expt}}

\vspace{0.5cm}

\centering{}\centering(a){}%
\begin{tabular}{ccc}
\hline 
E(eV)  & I (rel.)  & Electronic State\tabularnewline
\hline 
$1.54$  & $0.025$  & \multicolumn{1}{c}{$S_{1}$}\tabularnewline
$1.69$  & $0.011$  & \tabularnewline
$1.73$  & $0.016$  & \tabularnewline
$1.86$  & $0.007$  & \tabularnewline
$2.40$  & $0.02$  & $S_{2}$\tabularnewline
$2.54$  & $0.05$  & \multicolumn{1}{c}{$S_{3}$}\tabularnewline
$2.72$  & $0.05$  & \tabularnewline
$3.16$  & $0.08$  & $S_{4}$\tabularnewline
$3.29$  & $0.49$  & \multicolumn{1}{c}{$S_{5}$}\tabularnewline
$3.33$  & $0.43$  & \tabularnewline
$3.48$  & $0.46$  & $S_{6}$\tabularnewline
$3.68$  & $0.90$  & $S_{7}$\tabularnewline
$3.78$  & $1.00$  & \tabularnewline
\hline 
\end{tabular}\centering(b){}%
\begin{tabular}{ccc}
\hline 
E(eV)  & I (rel.)  & Electronic State\tabularnewline
\hline 
$1.43$  & $0.020$  & \multicolumn{1}{c}{$S_{1}$}\tabularnewline
$1.58$  & $0.006$  & \tabularnewline
$1.62$  & $0.016$  & \tabularnewline
$2.33$  & $0.033$  & $S_{3}$\tabularnewline
$2.50$  & $0.023$  & $S_{4}$\tabularnewline
$2.67$  & $0.030$  & $S_{5}$\tabularnewline
$2.80$  & $0.048$  & $S_{6}$\tabularnewline
$2.97$  & $0.19$  & \multicolumn{1}{c}{$S_{7}$}\tabularnewline
$3.14$  & $0.52$  & \tabularnewline
$3.42$  & $0.14$  & $S_{8}$\tabularnewline
$3.66$  & $1.00$  & $S_{9}$\tabularnewline
\hline 
\end{tabular}
\end{table}

\par\end{center}

\section{{\normalsize Conclusions }}

To summarize, we presented large-scale MRSDCI calculations on the
electronic structure and optical properties of oligoacenes, with focus
on the longer acenes, namely, octacene, nonacene, and decacene. By
performing such calculations on the lowest singlet and triplet states
of oligomers ranging from naphthalene up to decacene, we established
that the ground state in oligoacenes has singlet multiplicity, with
a singlet-triplet gap of approximately $0.5$ eV even for decacene.
The trends visible from our calculations rule out a singlet-triplet
crossover for the ground states of longer oligoacenes as well. This
result of ours has thus resolved an old speculation predicting that
nonacene onwards, the ground state of oligoacenes will be of triplet
multiplicity.\cite{Angliker}

Moreover, the many-body wavefunction analysis of the correlated singlet
ground state $1^{1}A_{g}^{-}$ reveals increasing contribution of
configurations with two open shells, accompanied with the decreasing
one from the closed-shell Hartree-Fock reference state, with the increasing
chain length. Thus our calculations predict an open-shell diradical
character for the singlet ground state of longer acenes.\cite{Bendikov}

As far as the singlet linear optical absorption is concerned, in all
the acenes, the first peak is due to a $y$-polarized transition to
the $1^{1}B_{2u}^{+}$ state, corresponding to the HOMO to LUMO transition.
The most intense state is the $x$-polarized transition to a $^{1}B_{3u}$
state, which is also dominated by single excitations. When we compare
our singlet linear absorption spectra of octacene and nonacene with
the experimental ones,\cite{Tonshoff2010} excellent agreement is
obtained on the important peak locations and intensity profiles. Furthermore,
the measured absorption spectra of longer acenes show no resemblance
with our computed triplet absorption spectra, confirming once again
the conclusion that the ground state of the longer acenes is indeed
singlet in nature.

Our calculations on the one-photon triplet absorption spectra predict
two intense $x$-polarized absorptions, which are well separated in
energy. Besides these two, there are a number of weaker peaks which
are either $x$ or $y$ polarized. This is in sharp contrast to the
singlet absorption which predicts only one intense peak. The existence
of two well separated $x$-polarized peaks in the triplet absorption
spectrum is one of the most important predictions of this work, and
can be tested in future experiments on oriented samples of longer
acenes.

We also performed singlet and triplet optical absorption calculations
on decacene, a molecule which has not been synthesized yet. We are
hopeful that in future, once decacene is synthesized in the laboratory,
our theoretical predictions could be tested in experiments.

In this paper we restricted ourselves to the low-lying excited states
of longer acenes which contribute to their linear optical properties.
However, not many calculations have been performed as far as the nonlinear
optical properties of these materials are concerned. In particular,
it will be of interest to compute the nonlinear susceptibilities corresponding
to two-photon absorption, and third harmonic generation. Both these
nonlinear optical processes have the capability to probe the higher
excited states of polyacenes, which is essential in order to obtain
a deeper understanding of the optical response of $\pi$ electrons.
At present, studies along these directions are underway in our group.

\acknowledgement

The authors thank Professor H. F. Bettinger (University of Tübingen)
for communicating his experimental data before publications. One of
us (H. C.) acknowledges the Council of Scientific and Industrial Research,
(CSIR), India for the financial support (SRF award No. 20-06/2009(i)EU-IV).

\begin{suppinfo}

\section*{Convergence of Excitation Energies in MRSDCI Calculations}

\begin{figure}[H]
\caption{Behavior of $1^{1}B_{2u}$ (circles), $1^{3}B_{2u}$ (squares) excited
states of nonacene with respect to the number of reference configurations
($N_{ref}$) included in the MRSDCI calcualtions performed using the
screened parameters.\label{fig:convergence}}

\vspace{0.5cm}

\includegraphics[scale=0.3]{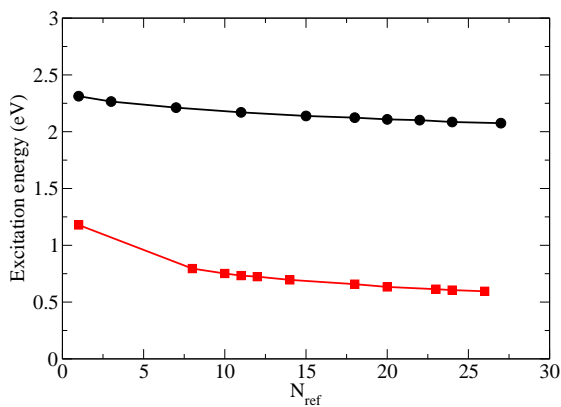}
\end{figure}

\textcolor{black}{In order to demonstate the convergence of our MRSDCI
calculations, we present the plot of the excitation energies of two
lowest states, singlet $1^{1}B_{2u}^{+}$ and triplet $1^{3}B_{2u}^{+}$,
of nonacene, computed using the screened parameters, in Figure \ref{fig:convergence},
calculated with increasing number of reference configurations ($N_{ref})$.
It is obvious from the figure that the convergence has been achieved
by the time twenty most important configurations ($N_{ref}=20)$ have
been included in the calculations.}

\section*{\textcolor{black}{Influence of the Geometry on the Optical Absorption
in Long Acenes}}

\textcolor{black}{The issue of the ground state geometry in long acenes
has been in debate for many years.\cite{Ramasesha-instability-12002,Ramasesha-acene-2002,Bendikov,Hachmann2007,FPA2011}
Some theoretical calculations have predicted a symmetric ground state
geometry to be lower,\cite{Ramasesha-instability-12002,Ramasesha-acene-2002}
while others have indicated a highly non-uniform geometry to be the
true ground state.\cite{Bendikov,Hachmann2007,FPA2011} In this work,
consistent with the PPP model based work of Ramasesha and coworkers,\cite{Ramasesha-acene-2002,Ramasesha-instability-12002}
and our own work,\cite{sony-acene-lo,sony-acene-pa} we have used
the symmetric ground state geometry for all oligoacenes, with all
the C-C bonds equal to $1.4$ \AA, and all bond angles taken to be
120$^{o}$\textsuperscript{}. In order to investigate the influence
of geometry on the optical absorption spectra, we performed calculations
on nonacene using a highly non-uniform geometry reported for its closed-shell
singlet ground state by Bendikov }\textcolor{black}{\emph{et al}}\textcolor{black}{.,\cite{Bendikov}
obtained using the B3LYP exchange-correlation functional in DFT. As
compared to the symmetric geometry, the nonuniformity is quite severe
in this geometry, with the two polyene chains exhibiting bond alternation,
and the interchain separation also varying significantly. The smallest
C-C bond length in the nonuniform structure is $\approx1.36$ Å, while
the largest one is close to $1.47$ Å. For this nonuniform geometry,
the hopping matrix elements between nearest-neighbor sites $i$ and
$j$, needed for the PPP calculations, were generated using the exponential
formula $t_{ij}=t_{0}e^{(r_{0}-r_{ij})/\delta}$, where $r_{ij}$
is the bond distance (in Å) between the sites, $t_{0}=-2.4$ eV, $r_{0}=1.4$
Å, and the decay constant $\delta=0.73$ Å. The value of $\delta$
was chosen so that the formula closely reproduces the hopping matrix
elements for a bond-alternating polyene with short/long bonds $1.35/1.45$
Å. The results of an SCI level calculation of the singlet optical
absorption spectrum of nonacene, both for the uniform (symmetric)
and this nonuniform geometry of Bendikov }\textcolor{black}{\emph{et
al}}\textcolor{black}{., \cite{Bendikov} computed using the screened
parameters of the PPP model, are presented in Figure. \ref{fig:role_of_geometry}.
From the figure it is obvious that there are small quantitative differences
between the two results, as far as peak locations are concerned. For
example, the first peak ($1^{1}B_{2u}$) is slightly blueshifted (0.11
eV) for the nonuniform geometry, as compared to the symmetric geometry,
while the most intense peak ($1^{1}B_{3u}$) is slightly redshifted
(0.15 eV). The only qualitative change between the two results is
at an energy higher than 5.5 eV, where one peak in the nonuniform
geometry is more intense as compared to its neighboring peak. Therefore,
we conclude that the variations in the ground state geometries of
the magnitude considered here, lead to small quantitative, and insignificant
qualitative, changes in the optical absorption spectra of oligoacenes.}
\begin{figure}[H]
\textcolor{black}{\caption{\textcolor{black}{Comparison of the calculated singlet optical absorption
spectra for nonacene, with the uniform (symmetric) geometry, and the
nonuniform geometry of Bendikov }\textcolor{black}{\emph{et al}}\textcolor{black}{.,
\cite{Bendikov} computed at SCI level, employing the screened parameters
in the PPP model.}\label{fig:role_of_geometry}}
}

\textcolor{black}{\vspace{0.8cm}
}

\textcolor{black}{\includegraphics[scale=0.3]{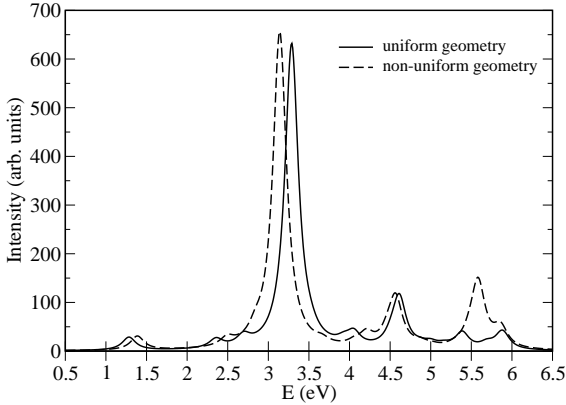}}
\end{figure}

\section*{Singlet linear absorption}

In the following we present a detailed description of the calculated
singlet linear absorption spectra of octacene, nonacene, and decacene,
presented in Figures \ref{fig:acene8_lo}-\ref{fig:acene10_lo} of
the main text for standard and the screened parameters.  
\begin{enumerate}
\item For all the oligoacene, the first peak is due to a $y$-polarized
transition, to the $1{}^{1}B_{2u}^{+}$ excited state of the system,
whose wave function is dominated by the $\vert H\rightarrow L\rangle$
single excitation, irrespective of the choice of the Coulomb parameters
employed in the PPP model. 
\item The second peak also corresponds to a $y$-polarized transition, to
the $2^{1}B_{2u}^{+}$ excited state of the system. The most important
configuration contributing to the many-particle wave function of this
state is $\vert H-1\rightarrow L+1\rangle$ excitation, irrespective
of the choice of the Coulomb parameters. 
\item The nature of the third peak is dependent upon the Coulomb parameters
employed in the PPP model. For the standard parameter case, this peak
always corresponds to the $x$-polarized, $1^{1}B_{3u}^{+}$ excited
state, signalling the onset of the most intense absorption feature
in the system. For octacene, it is the single most intense peak, whose
wave function mainly consists of single excitations $\vert H\rightarrow L+3\rangle+c.c.$
($c.c.$ denotes the charge conjugated configuration). For nonacene
and decacene, perhaps due to band formation, it is the first of the
two adjacent intense peaks which are $x$ polarized. For nonacene
it is the most intense peak, and the corresponding many-body wave
function is dominated by the single excitation $\vert H\rightarrow L+4\rangle+c.c.$.
For decacene, however, it is the second most intense feature, and
the double excitations $|H\rightarrow L;\, H\rightarrow L+1\rangle+c.c.$
contribute the most to its many-body wave function. \\
 \\
 In the screened parameter calculations, the third peak is a faint
peak, containing a mixture of $x$ and $y$ polarized transition to
the states, $1^{1}B_{3u}^{+}$ and $3^{1}B_{2u}^{+}$ of the system.
For all the oligoacenes, the double excitation $\vert H\rightarrow L;\, H\rightarrow L+1\rangle+c.c.$
contributes the most to the $1^{1}B_{3u}^{+}$ state, and the single
excitation $\vert H\rightarrow L+2\rangle+c.c.$ dominates the $3^{1}B_{2u}^{+}$
wave function. 
\item For the standard parameter case, the fourth peak is a faint peak for
octacene, consisting of a mixture of $x$- and $y$-polarized transitions
to the states, $2{}^{1}B_{3u}^{+}$ and $4^{1}B_{2u}^{+}$. Double
excitations $\vert H\rightarrow L;\, H\rightarrow L+1\rangle+c.c.$
contribute the most to the $2{}^{1}B_{3u}^{+}$ state, and single
excitations $\vert H-2\rightarrow L+2\rangle+c.c.$ to the $4^{1}B_{2u}^{+}$
state. For nonacene and decacene, however, it is an intense $x$-polarized
feature corresponding to the state $2{}^{1}B_{3u}^{+}$, which is
adjacent to their $1^{1}B_{3u}^{+}$ state mentioned above. For the
case of decacene, it also the most intense absorption of the system.
Both for nonacene and decacene, the single excitations $\vert H\rightarrow L+4\rangle+c.c.$
contribute the most to the many-particle wave function of this state.\\
 \\
 For the screened parameter calculations, the fourth peak corresponds
to the most intense absorption, through an $x$-polarized photon,
to the $2{}^{1}B_{3u}^{+}$ state, in case of octacene and nonacene.
For decacene, however, it appears as a shoulder to the most intense
peak, consisting of a mixture of $x$ and $y$ polarized transitions,
to the $2{}^{1}B_{3u}^{+}$ and $4^{1}B_{2u}^{+}$ states. The many-particle
wave function of the $2{}^{1}B_{3u}^{+}$ state for octacene and nonacene
is dominated by the single excitations $\vert H\rightarrow L+4\rangle+c.c.$,
while for decacene the double excitations $\vert H\rightarrow L+1;\, H\rightarrow L+2\rangle+c.c.$
dominate. The $4^{1}B_{2u}^{+}$ state of decacene has the maximum
contribution from the single excitation $\vert H-2\rightarrow L+2\rangle$. 
\item For the standard parameter case, the fifth peak is a very faint feature
consisting of $y$-polarized transition to the state $7^{1}B_{2u}^{+}$
for octacene, whose wave function is dominated by the single excitation
$\vert H-2\rightarrow L+2\rangle$. However, for nonacene and decacene,
it corresponds to the mixture of $x$ and $y$ polarized transitions
to the states, $3{}^{1}B_{3u}^{+}$ and $5^{1}B_{2u}^{+}$, respectively.
The double excitations $\vert H\rightarrow L+1;\, H-1\rightarrow L+1\rangle+c.c.$
for the $3{}^{1}B_{3u}^{+}$ state, and the triple excitation $\vert H\rightarrow L;\, H\rightarrow L;\, H-1\rightarrow L+1\rangle$
for the $5{}^{1}B_{2u}^{+}$ state, contribute the most to their wave
functions. \\
 \\
 For the screened parameter case, it is a faint peak consisting of
a mixture of $x$ and $y$ polarized transitions to the states, $3{}^{1}B_{3u}^{+}$
and $4{}^{1}B_{2u}^{+}$, respectively, for both octacene and nonacene.
For decacene, however, it is the most intense peak corresponding to
$x$ polarized transition to the state $4{}^{1}B_{3u}^{+}$. The most
important configuration contributing to the many-particle wave function
of the state for octacene and nonacene, is a mixture of double excitations
$\vert H\rightarrow L+1;\, H\rightarrow L+2\rangle+c.c.$ for $3{}^{1}B_{3u}^{+}$
state, and single excitations $\vert H-2\rightarrow L+2\rangle$ for
the $4{}^{1}B_{2u}^{+}$ state. For decacene, the single excitations
$\vert H\rightarrow L+5\rangle+c.c.$ contribute the most to the wavefunction
of this state. 
\item For the standard parameter case, the sixth peak corresponds to a mixture
of $x$ and $y$ polarized transitions to the states, $4{}^{1}B_{3u}^{+}$
and $9^{1}B_{2u}^{+}$, respectively, for octacene. The many-body
wave functions of both these states derive predominant contributions
from double excitations: $\vert H\rightarrow L;\, H-1\rightarrow L+2\rangle+c.c.$
for the $4{}^{1}B_{3u}^{+}$, and $\vert H\rightarrow L+1;\, H\rightarrow L+3\rangle+c.c.$
for the $9{}^{1}B_{2u}^{+}$ state. For nonacene (decacene), it corresponds
to a $y$-polarized transition to the state $10^{1}B_{2u}^{+}$ ($9{}^{1}B_{2u}^{+}$),
whose many-body wave function derives maximum contribution from the
single excitation $\vert H-2\rightarrow L+2\rangle$. \\
 \\
 The same peak with the screened parameters is a mixture of $x$ and
$y$ polarized transitions for the case of octacene and nonacene.
For octacene, transitions to states $6{}^{1}B_{3u}^{+}$ and $8{}^{1}B_{2u}^{+}$
constitute this peak, with double excitations $\vert H\rightarrow L;\, H\rightarrow L+1\rangle+c.c.$
and $\vert H\rightarrow L;\, H\rightarrow L+5\rangle+c.c.$, respectively,
contributing most to their wave functions. For nonacene, this peak
consists of three states $4{}^{1}B_{3u}^{+}$ and $6{}^{1}B_{2u}^{+}$
and $7{}^{1}B_{2u}^{+}$, with double excitation $\vert H\rightarrow L;\, H-1\rightarrow L+2\rangle+c.c.$
contributing the most to the $4{}^{1}B_{3u}^{+}$, triple excitation
$\vert H\rightarrow L;\, H\rightarrow L;\, H-1\rightarrow L+1\rangle$
to the $6{}^{1}B_{2u}^{+}$, and the single excitation $\vert H-2\rightarrow L+2\rangle$
to the $7{}^{1}B_{2u}^{+}$ state. However, for decacene, the peak
is purely $x$ polarized, due to transition to the state $6{}^{1}B_{3u}^{+}$
whose wave function derives maximum contribution from the double excitations
$\vert H\rightarrow L;\, H\rightarrow L+1\rangle+c.c.$. 
\item The seventh peak in the standard parameter spectrum is both $x$ and
$y$ polarized for all the three oligomers. For octacene, the peak
involves $5{}^{1}B_{3u}^{+}$ and $10^{1}B_{2u}^{+}$ states, with
single excitations $\vert H\rightarrow L+7\rangle+c.c.$ and $\vert H-2\rightarrow L+2\rangle$,
respectively, contributing the most to their wave functions. For nonacene
states $7{}^{1}B_{3u}^{+}$ and $12^{1}B_{2u}^{+}$ constitute the
peak, with single excitation $\vert H\rightarrow L+8\rangle$+c.c.
providing the maximum contribution to the $7{}^{1}B_{3u}^{+}$ state,
and triple excitation $\vert H\rightarrow L;\, H-1\rightarrow L+1;\, H-1\rightarrow L+1\rangle$
to the $12{}^{1}B_{2u}^{+}$ state. For decacene, the peak involves
$6{}^{1}B_{3u}^{+}$ and $11^{1}B_{2u}^{+}$ states, with double excitations
$\vert H\rightarrow L;\, H-1\rightarrow L+2\rangle+c.c.$ and triple
excitations $\vert H\rightarrow L;\, H-1\rightarrow L+1;\, H-1\rightarrow L+1\rangle$,
respectively, contributing the most to their wave functions. \\
 \\
 With the screened parameters, the seventh peak of octacene and nonacene
is $x$ polarized, while for decacene it has mixed $x$ and $y$ polarizations.
Both for octacene and nonacene states $8^{1}B_{3u}^{+}$ constitute
this peak, with single excitation $\vert H-1\rightarrow L+5\rangle+c.c.$
and $\vert H-1\rightarrow L+6\rangle+c.c.$ providing main contributions
to the wave functions of octacene, and nonacene, respectively. However,
for decacene, transitions to states $8{}^{1}B_{3u}^{+}$ and $10^{1}B_{2u}^{+}$
form the peak, with double excitations $\vert H\rightarrow L;\, H-1\rightarrow L+4\rangle+c.c.$
and single excitations $\vert H-1\rightarrow L+7\rangle+c.c.$, respectively,
contributing the most to their wave functions. 
\item With the standard parameters, the eighth peak does not exist for octacene,
while for nonacene and decacene, it has mixed $x$ and $y$ polarizations.
For nonacene, the states constituting the peak are $8{}^{1}B_{3u}^{+}$
and $16^{1}B_{2u}^{+}$, with single excitations $\vert H-1\rightarrow L+5\rangle+c.c.$
and triple excitations $\vert H\rightarrow L;\, H\rightarrow L+2;\, H-1\rightarrow L+1\rangle+c.c.$,
respectively, contributing to their wave functions. For decacene,
three states $7{}^{1}B_{3u}^{+}$, $14^{1}B_{2u}^{+}$, and $15^{1}B_{2u}^{+}$,
form this peak, with single excitations $\vert H\rightarrow L+8\rangle+c.c.$,
$\vert H\rightarrow L+12\rangle+c.c.$, and $\vert H\rightarrow L+9\rangle+c.c.$,
respectively, contributing the most to their many-particle wave functions.
\\
 \\
 With the screened parameters case, the eighth peak is $x$ and $y$
polarized for octacene and decacene, while it is only $x$ polarized
for nonacene. For octacene, states $9{}^{1}B_{3u}^{+}$ and $13{}^{1}B_{2u}^{+}$
form this peak, whose wave functions, respectively, derive maximum
contributions from single excitations $\vert H-2\rightarrow L+4\rangle+c.c.$,
and triple excitations $\vert H\rightarrow L;\, H-1\rightarrow L+1;\, H-1\rightarrow L+1\rangle$.
In case of nonacene, the transition to the state $9{}^{1}B_{3u}^{+}$
leads to this peak, with single excitations $\vert H-2\rightarrow L+4\rangle+c.c.$
contributing the most to its many-particle wave function. For decacene,
states $9^{1}B_{3u}^{+}$ and $14^{1}B_{2u}^{+}$ constitute the peak,
with the single excitations $\vert H\rightarrow L+8\rangle+c.c.$
and $\vert H\rightarrow L+9\rangle+c.c.$, respectively, providing
the largest contributions to their wave functions. 
\item The ninth peak, for the standard parameter case, exists only for decacene,
and has a mixed $x$ and $y$ polarization, with states $8{}^{1}B_{3u}^{+}$
and $17{}^{1}B_{2u}^{+}$ forming the peak. The single excitations
$\vert H-2\rightarrow L+4\rangle+c.c.$ contribute the most to $8{}^{1}B_{3u}^{+}$
state, while double excitations $\vert H\rightarrow L;\, H-1\rightarrow L+4\rangle+c.c.$
to the $17{}^{1}B_{2u}^{+}$ state. \\
 \\
 With the screened parameters, the ninth peak does not exist for nonacene,
and it is $x$ and $y$ polarized for octacene and decacene. For octacene,
states $12{}^{1}B_{3u}^{+}$ and $16{}^{1}B_{2u}^{+}$ constitute
this peak, with the double excitations $\vert H\rightarrow L;\, H-1\rightarrow L+6\rangle+c.c.$
and single excitations $\vert H-1\rightarrow L+7\rangle+c.c.$, respectively,
contributing the most to their wave functions. For decacene, the states
$11^{1}B_{3u}^{+}$ and $16^{1}B_{2u}^{+}$ form this peak, with the
single excitations $\vert H-2\rightarrow L+5\rangle+c.c.$ and $\vert H-1\rightarrow L+7\rangle+c.c.$,
respectively, providing the maximum contributions to their wave functions. 
\end{enumerate}
\begin{center}
\begin{table}
\caption{Excited states contributing to the singlet linear absorption spectrum
of octacene computed using the MRSDCI method coupled with the standard
parameters in the PPP model Hamiltonian. The table includes many-particle
dominant contributing configurations, excitation energies, dipole
matrix elements, and relative oscillator strengths (ROS) of various
states with respect to the $1^{1}A_{g}$ ground state. DF corresponds
to dipole forbidden state. Below, `$+c.c.$' indicates that the coefficient
of charge conjugate of a given configuration has the same sign, while
`$-c.c.$' implies that the two coefficients have opposite signs.\label{tab:acene-8} }

\centering{}%
\begin{tabular}{cccccc}
\hline 
Peak  & State  & E (eV)  & Transition  & ROS  & dominant contributing configurations\tabularnewline
\hline 
 &  &  & Dipole (\AA{})  &  & \tabularnewline
\hline 
DF  & $1^{1}B_{3u}^{-}$  & 2.31  & 0  & 0  & $|H\rightarrow L;H\rightarrow L+1\left\rangle \right.-c.c.(0.4948)$\tabularnewline
 &  &  &  &  & $|H\rightarrow L;H-1\rightarrow L+2\left\rangle \right.-c.c.(0.1938)$\tabularnewline
I  & $1^{1}B_{2u}^{+}$  & 2.24  & 0.905  & 0.034  & $|H\rightarrow L\left\rangle \right.(-0.8471)$\tabularnewline
 &  &  &  &  & $|H\rightarrow L;H\rightarrow L;H-1\rightarrow L+1\left\rangle \right.(0.0948)$\tabularnewline
II  & $2^{1}B_{2u}^{+}$  & 3.34  & 0.641  & 0.025  & $|H-1\rightarrow L+1\left\rangle \right.(-0.5826)$\tabularnewline
 &  &  &  &  & $|H\rightarrow L+2\left\rangle \right.-c.c.(0.4284)$\tabularnewline
III  & $1^{1}B_{3u}^{+}$  & 4.17  & 3.622  & 1.000  & $|H\rightarrow L+3\left\rangle \right.+c.c.(0.4622)$\tabularnewline
 &  &  &  &  & $|H\rightarrow L;H\rightarrow L+1\left\rangle \right.+c.c.(0.2614)$\tabularnewline
IV  & $4^{1}B_{2u}^{+}$  & 4.51  & 0.367  & 0.011  & $|H-2\rightarrow L+2\left\rangle \right.(-0.4148)$\tabularnewline
 &  &  &  &  & $|H-1\rightarrow L+4\left\rangle \right.+c.c.(0.4020)$\tabularnewline
 & $2^{1}B_{3u}^{+}$  & 4.57  & 1.079  & 0.097  & $|H\rightarrow L;H\rightarrow L+1\left\rangle \right.+c.c.(0.3214)$\tabularnewline
 &  &  &  &  & $|H\rightarrow L+3\left\rangle \right.+c.c.(0.3214)$\tabularnewline
V  & $7^{1}B_{2u}^{+}$  & 5.46  & 0.375  & 0.014  & $|H-2\rightarrow L+2\left\rangle \right.(0.3574)$\tabularnewline
 &  &  &  &  & $|H\rightarrow L+6\left\rangle \right.-c.c.(0.3489)$\tabularnewline
VI  & $9^{1}B_{2u}^{+}$  & 5.64  & 0.261  & 0.007  & $|H\rightarrow L+1;H\rightarrow L+3\left\rangle \right.+c.c.(0.3883)$\tabularnewline
 &  &  &  &  & $|H\rightarrow L;H\rightarrow L+5\left\rangle \right.+c.c.(0.2768)$\tabularnewline
 & $4^{1}B_{3u}^{+}$  & 5.65  & 0.335  & 0.012  & $|H\rightarrow L;H-1\rightarrow L+2\left\rangle \right.-c.c.(0.4696)$\tabularnewline
 &  &  &  &  & $|H\rightarrow L;H\rightarrow L+1\left\rangle \right.+c.c.(0.1289)$\tabularnewline
VII  & $10{}^{1}B_{2u}^{+}$  & 5.89  & 0.203  & 0.004  & $|H-2\rightarrow L+2\left\rangle \right.(-0.3657)$\tabularnewline
 &  &  &  &  & $|H-3\rightarrow L+3\left\rangle \right.(0.3074)$\tabularnewline
 & $5^{1}B_{3u}^{+}$  & 5.94  & 0.307  & 0.010  & $|H\rightarrow L+7\left\rangle \right.-c.c.(0.3397)$\tabularnewline
 &  &  &  &  & $|H\rightarrow L;H\rightarrow L+4\left\rangle \right.+c.c.(0.2494)$\tabularnewline
\hline 
\end{tabular}
\end{table}

\par\end{center}

\begin{center}
\begin{table}
\caption{Excited states contributing to the singlet linear absorption spectrum
of octacene computed using the MRSDCI method coupled with the screened
parameters in the PPP model Hamiltonian. The rest of the information
is same as that in Table \ref{tab:acene-8}.\label{tab:acene-8-scr}}

\centering{}%
\begin{tabular}{cccccc}
\hline 
Peak  & State  & E (eV)  & Transition  & ROS  & dominant contributing configurations\tabularnewline
\hline 
 &  &  & Dipole (\AA{})  &  & \tabularnewline
\hline 
DF  & $1^{1}B_{3u}^{-}$  & 1.59  & 0  & 0  & $|H\rightarrow L;H\rightarrow L+1\left\rangle \right.-c.c.(0.5078)$\tabularnewline
 &  &  &  &  & $|H\rightarrow L;H-1\rightarrow L+2\left\rangle \right.-c.c.(0.1949)$\tabularnewline
I  & $1^{1}B_{2u}^{+}$  & 1.49  & 1.241  & 0.050  & $|H\rightarrow L\left\rangle \right.(0.8503)$\tabularnewline
 &  &  &  &  & $|H\rightarrow L;H\rightarrow L;H-1\rightarrow L+1\left\rangle \right.(0.0929)$\tabularnewline
II  & $2^{1}B_{2u}^{+}$  & 2.65  & 0.897  & 0.047  & $|H-1\rightarrow L+1\left\rangle \right.(-0.7244)$\tabularnewline
 &  &  &  &  & $|H\rightarrow L+2\left\rangle \right.+c.c.(0.2638)$\tabularnewline
III  & $3^{1}B_{2u}^{+}$  & 2.89  & 0.440  & 0.012  & $|H\rightarrow L+2\left\rangle \right.+c.c.(0.5326)$\tabularnewline
 &  &  &  &  & $|H-1\rightarrow L+1\left\rangle \right.(-0.3290)$\tabularnewline
 & $1^{1}B_{3u}^{+}$  & 2.97  & 0.845  & 0.046  & $|H\rightarrow L;H\rightarrow L+1\left\rangle \right.-c.c.(0.4942)$\tabularnewline
 &  &  &  &  & $|H\rightarrow L+1;H-1\rightarrow L+1\left\rangle \right.-c.c.(0.2522)$\tabularnewline
IV  & $2^{1}B_{3u}^{+}$  & 3.38  & 3.675  & 1.000  & $|H\rightarrow L+4\left\rangle \right.+c.c.(0.5831)$\tabularnewline
 &  &  &  &  & $|H-1\rightarrow L+5\left\rangle \right.+c.c.(0.1004)$\tabularnewline
V  & $3^{1}B_{3u}^{+}$  & 3.91  & 0.410  & 0.014  & $|H\rightarrow L+1;H\rightarrow L+2\left\rangle \right.-c.c.(0.4028)$\tabularnewline
 &  &  &  &  & $|H\rightarrow L+1;H\rightarrow L+4\left\rangle \right.-c.c.(0.3003)$\tabularnewline
 & $4^{1}B_{2u}^{+}$  & 3.97  & 0.641  & 0.036  & $|H-2\rightarrow L+2\left\rangle \right.(-0.5280)$\tabularnewline
 &  &  &  &  & $|H-1\rightarrow L+3\left\rangle \right.+c.c.(0.3722)$\tabularnewline
VI  & $8^{1}B_{2u}^{+}$  & 4.52  & 0.230  & 0.005  & $|H\rightarrow L;H\rightarrow L+5\left\rangle \right.-c.c.(0.4706)$\tabularnewline
 &  &  &  &  & $|H\rightarrow L;H-1\rightarrow L+4\left\rangle \right.-c.c.(0.3075)$\tabularnewline
 & $6^{1}B_{3u}^{+}$  & 4.57  & 0.403  & 0.016  & $|H\rightarrow L;H\rightarrow L+1\left\rangle \right.-c.c.(0.3419)$\tabularnewline
 &  &  &  &  & $|H\rightarrow L;H\rightarrow L+3\left\rangle \right.-c.c.(0.2890)$\tabularnewline
VII  & $8{}^{1}B_{3u}^{+}$  & 4.80  & 1.301  & 0.178  & $|H-1\rightarrow L+5\left\rangle \right.+c.c.(0.5537)$\tabularnewline
 &  &  &  &  & $|H\rightarrow L;H-1\rightarrow L+2\left\rangle \right.-c.c.(0.1321)$\tabularnewline
VIII  & $9{}^{1}B_{3u}^{+}$  & 5.14  & 0.410  & 0.019  & $|H-2\rightarrow L+4\left\rangle \right.+c.c.(0.5720)$\tabularnewline
 &  &  &  &  & $|H\rightarrow L+2;H\rightarrow L+3\left\rangle \right.-c.c.(0.0943)$\tabularnewline
 & $13{}^{1}B_{2u}^{+}$  & 5.22  & 0.418  & 0.020  & $|H\rightarrow L;H-1\rightarrow L+1;H-1\rightarrow L+1\left\rangle \right.(0.3505)$\tabularnewline
 &  &  &  &  & $|H-1\rightarrow L+7\left\rangle \right.+c.c.(0.2709)$\tabularnewline
IX  & $16{}^{1}B_{2u}^{+}$  & 5.53  & 0.440  & 0.023  & $|H-1\rightarrow L+7\left\rangle \right.+c.c.(0.3983)$\tabularnewline
 &  &  &  &  & $|H-3\rightarrow L+3\left\rangle \right.(0.3636)$\tabularnewline
 & $12{}^{1}B_{3u}^{+}$  & 5.66  & 0.154  & 0.003  & $|H\rightarrow L;H-1\rightarrow L+6\left\rangle \right.-c.c.(0.4014)$\tabularnewline
 &  &  &  &  & $|H\rightarrow L+2;H-2\rightarrow L+1\left\rangle \right.-c.c.(0.3019)$\tabularnewline
\hline 
\end{tabular}
\end{table}

\par\end{center}

\begin{center}
\begin{table}
\caption{Excited states contributing to the singlet linear absorption spectrum
of nonacene computed using the MRSDCI method coupled with the standard
parameters in the PPP model Hamiltonian. The rest of the information
is same as that in Table \ref{tab:acene-8}.\label{tab:acene-9}}

\centering{}%
\begin{tabular}{cccccc}
\hline 
Peak  & State  & E (eV)  & Transition  & ROS  & dominant contributing configurations\tabularnewline
\hline 
 &  &  & Dipole (\AA{})  &  & \tabularnewline
\hline 
DF  & $1^{1}B_{3u}^{-}$  & 1.86  & 0  & 0  & $|H\rightarrow L;H\rightarrow L+1\left\rangle \right.+c.c.(0.4872)$\tabularnewline
 &  &  &  &  & $|H\rightarrow L+1;H\rightarrow L+2\left\rangle \right.+c.c.(0.2027)$\tabularnewline
I  & $1^{1}B_{2u}^{+}$  & 1.82  & 1.328  & 0.092  & $|H\rightarrow L\left\rangle \right.(0.8290)$\tabularnewline
 &  &  &  &  & $|H\rightarrow L;H\rightarrow L;H-1\rightarrow L+1\left\rangle \right.(0.1066)$\tabularnewline
II  & $2^{1}B_{2u}^{+}$  & 2.79  & 0.733  & 0.043  & $|H-1\rightarrow L+1\left\rangle \right.(0.5506)$\tabularnewline
 &  &  &  &  & $|H\rightarrow L+2\left\rangle \right.+c.c.(0.4342)$\tabularnewline
III  & $1^{1}B_{3u}^{+}$  & 3.80  & 3.037  & 1.000  & $|H\rightarrow L+4\left\rangle \right.-c.c.(0.3757)$\tabularnewline
 &  &  &  &  & $|H\rightarrow L;H\rightarrow L+1\left\rangle \right.-c.c.(0.3274)$\tabularnewline
IV  & $2^{1}B_{3u}^{+}$  & 4.12  & 2.583  & 0.786  & $|H\rightarrow L+4\left\rangle \right.-c.c.(0.3943)$\tabularnewline
 &  &  &  &  & $|H\rightarrow L;H\rightarrow L+1\left\rangle \right.-c.c.(0.2791)$\tabularnewline
V  & $5^{1}B_{2u}^{+}$  & 4.62  & 0.439  & 0.025  & $|H\rightarrow L;H\rightarrow L;H-1\rightarrow L+1\left\rangle \right.(0.6215)$\tabularnewline
 &  &  &  &  & $|H-1\rightarrow L+1\left\rangle \right.(0.2357)$\tabularnewline
 & $3^{1}B_{3u}^{+}$  & 4.63  & 0.549  & 0.040  & $|H\rightarrow L+1;H-1\rightarrow L+1\left\rangle \right.-c.c.(0.3339)$\tabularnewline
 &  &  &  &  & $|H\rightarrow L+1;H\rightarrow L+2\left\rangle \right.-c.c.(0.3035)$\tabularnewline
VI  & $10{}^{1}B_{2u}^{+}$  & 5.29  & 0.457  & 0.032  & $|H-2\rightarrow L+2\left\rangle \right.(-0.3824)$\tabularnewline
 &  &  &  &  & $|H\rightarrow L+1;H\rightarrow L+4\left\rangle \right.+c.c.(0.2681)$\tabularnewline
VII  & $7^{1}B_{3u}^{+}$  & 5.71  & 0.277  & 0.013  & $|H\rightarrow L+8\left\rangle \right.-c.c.(0.3021)$\tabularnewline
 &  &  &  &  & $|H\rightarrow L;H\rightarrow L+3\left\rangle \right.+c.c.(0.2619)$\tabularnewline
 & $12{}^{1}B_{2u}^{+}$  & 5.72  & 0.318  & 0.017  & $|H\rightarrow L;H-1\rightarrow L+1;H-1\rightarrow L+1\left\rangle \right.(-0.3293)$\tabularnewline
 &  &  &  &  & $|H\rightarrow L;H-1\rightarrow L+4\left\rangle \right.-c.c.(0.2869)$\tabularnewline
VIII  & $16{}^{1}B_{2u}^{+}$  & 5.94  & 0.287  & 0.014  & $|H\rightarrow L;H\rightarrow L+2;H-1\rightarrow L+1\left\rangle \right.-c.c.(0.3742)$\tabularnewline
 &  &  &  &  & $|H\rightarrow L;H-1\rightarrow L+4\left\rangle \right.-c.c.(0.2500)$\tabularnewline
 & $8^{1}B_{3u}^{+}$  & 5.95  & 0.996  & 0.169  & $|H-1\rightarrow L+5\left\rangle \right.-c.c.(0.3932)$\tabularnewline
 &  &  &  &  & $|H-2\rightarrow L+4\left\rangle \right.-c.c.(0.2993)$\tabularnewline
\hline 
\end{tabular}
\end{table}

\par\end{center}

\begin{center}
\begin{table}
\caption{Excited states contributing to the singlet linear absorption spectrum
of nonacene computed using the MRSDCI method coupled with the screened
parameters in the PPP model Hamiltonian. The rest of the information
is same as that in Table \ref{tab:acene-8}.\label{tab:acene-9-scr}}

\centering{}%
\begin{tabular}{cccccc}
\hline 
Peak  & State  & E (eV)  & Transition  & ROS  & dominant contributing configurations\tabularnewline
\hline 
 &  &  & Dipole (\AA{})  &  & \tabularnewline
\hline 
DF  & $1^{1}B_{3u}^{-}$  & 1.51  & 0  & 0  & $|H\rightarrow L;H\rightarrow L+1\left\rangle \right.+c.c.(0.5143)$\tabularnewline
 &  &  &  &  & $|H\rightarrow L;H-1\rightarrow L+2\left\rangle \right.-c.c.(0.2089)$\tabularnewline
I  & $1^{1}B_{2u}^{+}$  & 1.46  & 1.316  & 0.051  & $|H\rightarrow L\left\rangle \right.(0.8551)$\tabularnewline
 &  &  &  &  & $|H\rightarrow L;H\rightarrow L;H-1\rightarrow L+1\left\rangle \right.(0.1017)$\tabularnewline
II  & $2^{1}B_{2u}^{+}$  & 2.45  & 0.935  & 0.043  & $|H-1\rightarrow L+1\left\rangle \right.(0.7260)$\tabularnewline
 &  &  &  &  & $|H\rightarrow L+2\left\rangle \right.-c.c.(0.2689)$\tabularnewline
III  & $1^{1}B_{3u}^{+}$  & 2.71  & 0.611  & 0.020  & $|H\rightarrow L;H\rightarrow L+1\left\rangle \right.-c.c.(0.4901)$\tabularnewline
 &  &  &  &  & $|H\rightarrow L+1;H-1\rightarrow L+1\left\rangle \right.-c.c.(0.2454)$\tabularnewline
 & $3^{1}B_{2u}^{+}$  & 2.77  & 0.507  & 0.014  & $|H\rightarrow L+2\left\rangle \right.-c.c.(0.5434)$\tabularnewline
 &  &  &  &  & $|H-1\rightarrow L+1\left\rangle \right.(-0.3462)$\tabularnewline
IV  & $2^{1}B_{3u}^{+}$  & 3.32  & 3.887  & 1.000  & $|H\rightarrow L+4\left\rangle \right.-c.c.(0.5689)$\tabularnewline
 &  &  &  &  & $|H-1\rightarrow L+6\left\rangle \right.+c.c.(0.1185)$\tabularnewline
V  & $3^{1}B_{3u}^{+}$  & 3.63  & 0.782  & 0.044  & $|H\rightarrow L+1;H\rightarrow L+2\left\rangle \right.+c.c.(0.3441)$\tabularnewline
 &  &  &  &  & $|H-1\rightarrow L+1;H\rightarrow L+1\left\rangle \right.-c.c.(0.3221)$\tabularnewline
 & $4^{1}B_{2u}^{+}$  & 3.69  & 0.559  & 0.023  & $|H-2\rightarrow L+2\left\rangle \right.(-0.4715)$\tabularnewline
 &  &  &  &  & $|H-1\rightarrow L+3\left\rangle \right.+c.c.(0.4395)$\tabularnewline
VI  & $4{}^{1}B_{3u}^{+}$  & 3.93  & 0.369  & 0.011  & $|H\rightarrow L;H-1\rightarrow L+2\left\rangle \right.-c.c.(0.4953)$\tabularnewline
 &  &  &  &  & $|H\rightarrow L;H\rightarrow L+3\left\rangle \right.-c.c.(0.2163)$\tabularnewline
 & $6^{1}B_{2u}^{+}$  & 4.01  & 0.690  & 0.038  & $|H\rightarrow L;H\rightarrow L;H-1\rightarrow L+1\left\rangle \right.(0.5954)$\tabularnewline
 &  &  &  &  & $|H\rightarrow L+5\left\rangle \right.+c.c.(0.3166)$\tabularnewline
 & $7{}^{1}B_{2u}^{+}$  & 4.12  & 0.692  & 0.039  & $|H-2\rightarrow L+2\left\rangle \right.(-0.5949)$\tabularnewline
 &  &  &  &  & $|H-1\rightarrow L+3\left\rangle \right.+c.c.(0.3994)$\tabularnewline
VII  & $8{}^{1}B_{3u}^{+}$  & 4.61  & 1.450  & 0.193  & $|H-1\rightarrow L+6\left\rangle \right.+c.c.(0.5262)$\tabularnewline
 &  &  &  &  & $|H\rightarrow L;H-1\rightarrow L+2\left\rangle \right.-c.c.(0.1473)$\tabularnewline
VIII  & $9{}^{1}B_{3u}^{+}$  & 4.97  & 0.338  & 0.011  & $|H-2\rightarrow L+4\left\rangle \right.+c.c.(0.5823)$\tabularnewline
 &  &  &  &  & $|H\rightarrow L+8\left\rangle \right.+c.c.(0.0699)$\tabularnewline
\hline 
\end{tabular}
\end{table}

\par\end{center}

\begin{center}
\begin{table}
\caption{Excited states contributing to the singlet linear absorption spectrum
of decacene computed using the MRSDCI method coupled with the standard
parameters in the PPP model Hamiltonian. The rest of the information
is same as that in Table \ref{tab:acene-8}.\label{tab:acene-10}}

\centering{}%
\begin{tabular}{cccccc}
\hline 
Peak  & State  & E (eV)  & Transition  & ROS  & dominant contributing configurations\tabularnewline
\hline 
 &  &  & Dipole (\AA{})  &  & \tabularnewline
\hline 
DF  & $1^{1}B_{3u}^{-}$  & 1.72  & 0  & 0  & $|H\rightarrow L;H\rightarrow L+1\left\rangle \right.-c.c.(0.4790)$\tabularnewline
 &  &  &  &  & $|H\rightarrow L+1;H\rightarrow L+2\left\rangle \right.+c.c.(0.2119)$\tabularnewline
I  & $1^{1}B_{2u}^{+}$  & 1.79  & 1.423  & 0.084  & $|H\rightarrow L\left\rangle \right.(-0.8203)$\tabularnewline
 &  &  &  &  & $|H-1\rightarrow L+1\left\rangle \right.(-0.1213)$\tabularnewline
II  & $2^{1}B_{2u}^{+}$  & 2.64  & 0.776  & 0.037  & $|H-1\rightarrow L+1\left\rangle \right.(-0.5607)$\tabularnewline
 &  &  &  &  & $|H\rightarrow L+2\left\rangle \right.-c.c.(0.4239)$\tabularnewline
III  & $1^{1}B_{3u}^{+}$  & 3.68  & 2.554  & 0.553  & $|H\rightarrow L;H\rightarrow L+1\left\rangle \right.+c.c.(0.3710)$\tabularnewline
 &  &  &  &  & $|H\rightarrow L+4\left\rangle \right.+c.c.(0.2865)$\tabularnewline
IV  & $2^{1}B_{3u}^{+}$  & 4.00  & 3.293  & 1.000  & $|H\rightarrow L+4\left\rangle \right.+c.c.(0.4451)$\tabularnewline
 &  &  &  &  & $|H\rightarrow L+1;H\rightarrow L+2\left\rangle \right.-c.c.(0.2322)$\tabularnewline
V  & $5^{1}B_{2u}^{+}$  & 4.40  & 0.505  & 0.026  & $|H\rightarrow L;H\rightarrow L;H-1\rightarrow L+1\left\rangle \right.(-0.6203)$\tabularnewline
 &  &  &  &  & $|H-1\rightarrow L+1\left\rangle \right.(-0.2380)$\tabularnewline
 & $3^{1}B_{3u}^{+}$  & 4.46  & 0.739  & 0.056  & $|H\rightarrow L+1;H-1\rightarrow L+1\left\rangle \right.+c.c.(0.3459)$\tabularnewline
 &  &  &  &  & $|H\rightarrow L+1;H\rightarrow L+2\left\rangle \right.-c.c.(0.2800)$\tabularnewline
VI  & $9^{1}B_{2u}^{+}$  & 5.05  & 0.392  & 0.018  & $|H-2\rightarrow L+2\left\rangle \right.(-0.3901)$\tabularnewline
 &  &  &  &  & $|H-1\rightarrow L+3\left\rangle \right.(-0.2482)$\tabularnewline
VII  & $6^{1}B_{3u}^{+}$  & 5.45  & 0.337  & 0.014  & $|H\rightarrow L;H-1\rightarrow L+2\left\rangle \right.-c.c.(0.3992)$\tabularnewline
 &  &  &  &  & $|H\rightarrow L+8\left\rangle \right.-c.c.(0.2341)$\tabularnewline
 & $11{}^{1}B_{2u}^{+}$  & 5.45  & 0.351  & 0.015  & $|H\rightarrow L;H-1\rightarrow L+1;H-1\rightarrow L+1\left\rangle \right.(-0.4912)$\tabularnewline
 &  &  &  &  & $|H\rightarrow L;H\rightarrow L;H-1\rightarrow L+1\left\rangle \right.(-0.2510)$\tabularnewline
VIII  & $7^{1}B_{3u}^{+}$  & 5.59  & 0.584  & 0.044  & $|H\rightarrow L+8\left\rangle \right.(0.3796)$\tabularnewline
 &  &  &  &  & $|H-1\rightarrow L+6\left\rangle \right.-c.c.(0.2866)$\tabularnewline
 & $14{}^{1}B_{2u}^{+}$  & 5.62  & 0.332  & 0.014  & $|H\rightarrow L+12\left\rangle \right.+c.c.(0.4160)$\tabularnewline
 &  &  &  &  & $|H-1\rightarrow L+7\left\rangle \right.+c.c.(0.2309)$\tabularnewline
 & $15{}^{1}B_{2u}^{+}$  & 5.63  & 0.326  & 0.014  & $|H\rightarrow L+9\left\rangle \right.+c.c.(0.3036)$\tabularnewline
 &  &  &  &  & $|H-1\rightarrow L+7\left\rangle \right.+c.c.(0.2945)$\tabularnewline
IX  & $8^{1}B_{3u}^{+}$  & 5.89  & 0.913  & 0.113  & $|H-2\rightarrow L+4\left\rangle \right.-c.c.(0.3596)$\tabularnewline
 &  &  &  &  & $|H-1\rightarrow L+6\left\rangle \right.-c.c.(0.2949)$\tabularnewline
 & $17{}^{1}B_{2u}^{+}$  & 5.89  & 0.149  & 0.003  & $|H\rightarrow L;H-1\rightarrow L+4\left\rangle \right.-c.c.(0.4687)$\tabularnewline
 &  &  &  &  & $|H\rightarrow L;H\rightarrow L;H-2\rightarrow L+2\left\rangle \right.(-0.1847)$\tabularnewline
\hline 
\end{tabular}
\end{table}

\par\end{center}

\begin{center}
\begin{table}
\caption{Excited states contributing to the singlet linear absorption spectrum
of decacene computed using the MRSDCI method coupled with the screened
parameters in the PPP model Hamiltonian. The rest of the information
is same as that in Table \ref{tab:acene-8}.\label{tab:acene-10-scr}}

\centering{}%
\begin{tabular}{cccccc}
\hline 
Peak  & State  & E (eV)  & Transition  & ROS  & dominant contributing configurations\tabularnewline
\hline 
 &  &  & Dipole (\AA{})  &  & \tabularnewline
\hline 
DF  & $1^{1}B_{3u}^{-}$  & 1.15  & 0  & 0  & $|H\rightarrow L;H\rightarrow L+1\left\rangle \right.-c.c.(0.4851)$\tabularnewline
 &  &  &  &  & $|H\rightarrow L+1;H-1\rightarrow L+2\left\rangle \right.-c.c.(0.2102)$\tabularnewline
I  & $1^{1}B_{2u}^{+}$  & 1.27  & 1.369  & 0.067  & $|H\rightarrow L\left\rangle \right.(-0.8338)$\tabularnewline
 &  &  &  &  & $|H\rightarrow L;H\rightarrow L;H-1\rightarrow L+1\left\rangle \right.(+0.1112)$\tabularnewline
II  & $2^{1}B_{2u}^{+}$  & 2.16  & 0.929  & 0.052  & $|H-1\rightarrow L+1\left\rangle \right.(+0.7110)$\tabularnewline
 &  &  &  &  & $|H\rightarrow L+2\left\rangle \right.+c.c.(0.2625)$\tabularnewline
III  & $3^{1}B_{2u}^{+}$  & 2.42  & 0.417  & 0.012  & $|H\rightarrow L+2\left\rangle \right.+c.c.(0.5220)$\tabularnewline
 &  &  &  &  & $|H-1\rightarrow L+1\left\rangle \right.(-0.3185)$\tabularnewline
 & $1^{1}B_{3u}^{+}$  & 2.42  & 0.295  & 0.006  & $|H\rightarrow L;H\rightarrow L+1\left\rangle \right.+c.c.(0.4762)$\tabularnewline
 &  &  &  &  & $|H-1\rightarrow L;H-1\rightarrow L+1\left\rangle \right.+c.c.(0.2590)$\tabularnewline
IV  & $2^{1}B_{3u}^{+}$  & 3.20  & 1.872  & 0.315  & $|H\rightarrow L+1;H\rightarrow L+2\left\rangle \right.+c.c.(0.3455)$\tabularnewline
 &  &  &  &  & $|H\rightarrow L+5\left\rangle \right.+c.c.(0.2733)$\tabularnewline
 & $4^{1}B_{2u}^{+}$  & 3.21  & 0.706  & 0.045  & $|H-2\rightarrow L+2\left\rangle \right.(0.5140)$\tabularnewline
 &  &  &  &  & $|H-1\rightarrow L+3\left\rangle \right.+c.c.(0.3975)$\tabularnewline
V  & $4^{1}B_{3u}^{+}$  & 3.35  & 3.259  & 1.000  & $|H\rightarrow L+5\left\rangle \right.+c.c.(0.4970)$\tabularnewline
 &  &  &  &  & $|H\rightarrow L;H-1\rightarrow L+2\left\rangle \right.-c.c.(0.2002)$\tabularnewline
VI  & $6^{1}B_{3u}^{+}$  & 3.81  & 0.273  & 0.008  & $|H\rightarrow L;H\rightarrow L+1\left\rangle \right.-c.c.(0.3703)$\tabularnewline
 &  &  &  &  & $|H\rightarrow L;H\rightarrow L+3\left\rangle \right.+c.c.(0.3689)$\tabularnewline
VII  & $8{}^{1}B_{3u}^{+}$  & 4.32  & 0.184  & 0.004  & $|H\rightarrow L;H-1\rightarrow L+4\left\rangle \right.-c.c.(0.2891)$\tabularnewline
 &  &  &  &  & $|H\rightarrow L+6;H-5\rightarrow L+2\left\rangle \right.-c.c.(0.2750)$\tabularnewline
 & $10^{1}B_{2u}^{+}$  & 4.33  & 0.478  & 0.028  & $|H-1\rightarrow L+7\left\rangle \right.-c.c.(0.2951)$\tabularnewline
 &  &  &  &  & $|H-2\rightarrow L+4\left\rangle \right.-c.c.(0.2793)$\tabularnewline
VIII  & $9{}^{1}B_{3u}^{+}$  & 4.51  & 0.330  & 0.014  & $|H\rightarrow L+8\left\rangle \right.+c.c.(0.4467)$\tabularnewline
 &  &  &  &  & $|H\rightarrow L+2;H-1\rightarrow L+2\left\rangle \right.+c.c.(0.2155)$\tabularnewline
 & $14^{1}B_{2u}^{+}$  & 4.56  & 0.376  & 0.018  & $|H\rightarrow L+9\left\rangle \right.+c.c.(0.4944)$\tabularnewline
 &  &  &  &  & $|H-3\rightarrow L+7\left\rangle \right.(-0.2478)$\tabularnewline
IX  & $11{}^{1}B_{3u}^{+}$  & 4.77  & 0.754  & 0.076  & $|H-2\rightarrow L+5\left\rangle \right.+c.c.(0.5543)$\tabularnewline
 &  &  &  &  & $|H\rightarrow L+3;H-1\rightarrow L+1\left\rangle \right.-c.c.(0.2076)$\tabularnewline
 & $16^{1}B_{2u}^{+}$  & 4.77  & 0.520  & 0.036  & $|H-1\rightarrow L+7\left\rangle \right.-c.c.(0.3795)$\tabularnewline
 &  &  &  &  & $|H-3\rightarrow L+7\left\rangle \right.(0.3359)$\tabularnewline
\hline 
\end{tabular}
\end{table}

\par\end{center}

\section*{Triplet absorption}

In the following we present a detailed description of the calculated
triplet absorption spectra of octacene, nonacene, and decacene, presented
in Figures \ref{fig:acene8_pa}-\ref{fig:acene10_pa} of the main
text for the standard and the screened parameters.  
\begin{enumerate}
\item The first peak always corresponds to the $1^{3}B_{1g}^{-}$ excited
state of the system, whose wave function is dominated by the single
excitations $\vert H\rightarrow L+1\rangle+c.c.$, irrespective of
the oligoacene in question, or the Coulomb parameters employed. 
\item The second peak corresponds to a $y$-polarized transition to the
$1^{3}A_{g}^{-}$ excited state, for all the oligoacenes, irrespective
of the Coulomb parameters employed. The most important configuration
contributing to the many-particle wave function of the $1^{3}A_{g}^{-}$
state, is the double excitation $\vert H\rightarrow L;\, H-1\rightarrow L+1\rangle$. 
\item The nature of the third peak is dependent upon the Coulomb parameters
employed in the PPP model. For the standard parameter case, this peak
corresponds to a mixture of $x$- and $y$-polarized the transitions
to states, $2{}^{3}B_{1g}^{-}$ and $3^{3}A_{g}^{-}$ for octacene
and nonacene. For octacene, the single excitations $\vert H\rightarrow L+4\rangle+c.c.$
for the $2{}^{3}B_{1g}^{-}$ state, and $\vert H-1\rightarrow L+3\rangle+c.c.$
for $3^{3}A_{g}^{-}$ state contribute the most to the respective
wave functions. For nonacene, the single excitations $\vert H\rightarrow L+3\rangle+c.c.$
for the $2{}^{3}B_{1g}^{-}$ state, and the double excitation $\vert H\rightarrow L;\, H-1\rightarrow L+1\rangle$
for the $3^{3}A_{g}^{-}$ states dominate the corresponding wave functions.
However, for decacene, the peak corresponds to an $x$ polarized transition
to the state $2{}^{3}B_{1g}^{-}$ , whose wave function is dominated
by single excitations, with the configurations $\vert H\rightarrow L+3\rangle+c.c.$
contributing the most. \\
 \\
 For the screened parameter case, for all the oligomers, the third
peak is due to an $x$-polarized transition to the state $3{}^{3}B_{1g}^{-}$
whose wave function derives the maximum contribution from the single
excitations $\vert H\rightarrow L+3\rangle+c.c.$. 
\item As far as the fourth peak is concerned, with standard parameters it
corresponds to an $x$-polarized transition to the $3{}^{3}B_{1g}^{-}$
excited state, for all the oligoacenes. For octacene, it happens to
be the most intense peak of the spectrum, but for nonacene and decacene,
it is a shoulder to the most intense peak. Double excitations $|H\rightarrow L;\, H\rightarrow L+3\rangle+c.c.$
contribute the most to the many-particle wave function of this state
for octacene, whereas for nonacene and decacene, the single excitations
$\vert H-1\rightarrow L+2\rangle+c.c.$ dominate the wave function.
\\
 \\
 In the screened parameter spectrum, the fourth peak is due to a $y$-polarized
transiton to the $3{}^{3}A_{g}^{-}$ state for octacene and nonacene.
For octacene, the single excitations $\vert H\rightarrow L+5\rangle+c.c.$
contribute the most to the many particle wave function of this state,
while for nonacene, the single excitations $\vert H\rightarrow L+6\rangle+c.c.$
dominate the state. However, for decacene, the fourth peak is the
second most intense peak of the spectrum, corresponding to an $x$-polarized
transition to the $6{}^{3}B_{1g}^{-}$ state, whose wave function
is dominated by the double excitations $\vert H\rightarrow L;\, H\rightarrow L+5\rangle+c.c.$. 
\item For the standard parameter case, the fifth peak is due to an $x$-polarized
transition to the state $4{}^{3}B_{1g}^{-}$ for all the oligoacenes.
For the case of octacene, it appears as a shoulder of the most intense
peak (peak IV), while for nonacene and decacene it is the most intense
peak. For octacene the wave function of this state is composed mainly
of single excitations, with configurations $|H-1\rightarrow L+2\rangle+c.c.$
contributing the most. For nonacene and decacene, however, the wave
function is dominated by the double excitations $\vert H\rightarrow L;\, H\rightarrow L+4\rangle+c.c.$.
\\
 \\
 In the screened parameter spectrum also the fifth peak corresponds
to an $x$ polarized transition for all the oligomers. For octacene,
nonacene, and decacene, the states involved are $4{}^{3}B_{1g}^{-}$,
$6{}^{3}B_{1g}^{-}$, and $8{}^{3}B_{1g}^{-}$, respectively. For
octacene and nonacene, this peak is the second most intense one of
the corresponding spectra, and the most important configuration contributing
to the many-particle wave function of the states are the double excitations
$\vert H\rightarrow L;\, H\rightarrow L+4\rangle+c.c.$. However,
for decacene, it is a relatively weaker feature, with the double excitations
$\vert H-1\rightarrow L;\, H\rightarrow L+6\rangle+c.c.$ dominating
the wave function of the state. 
\item The sixth peak in the standard parameter spectrum, is formed by a
$y$-polarized transition to the state $9{}^{3}A{}_{g}^{-}$ for the
case of octacene, whose wave function is dominated by the single excitations
$\vert H\rightarrow L+5\rangle+c.c.$ and $\vert H-3\rightarrow L+1\rangle+c.c.$
However, for nonacene and decacene, it corresponds to an $x$-polarized
transition to the state $6{}^{3}B{}_{1g}^{-}$, whose wave function
receives maximum contribution from the triple excitations $\vert H\rightarrow L+1;\, H\rightarrow L+1;\, H-1\rightarrow L\rangle+c.c.$.
\\
 \\
 With the screened parameters, this peak is formed by an $x$-polarized
transition to the state $7{}^{3}B_{1g}^{-}$ for octacene, with double
excitations $\vert H\rightarrow L+1;\, H-5\rightarrow L\rangle+c.c.$
dominating its wave function. For nonacene and decacene, the peak
is due to mixed $x$ and $y$ polarized transitions. For nonacene,
states $8^{3}B_{1g}^{-}$ and $10{}^{3}A{}_{g}^{-}$ form this peak,
with their wave functions dominated by double excitations $\vert H\rightarrow L+1;\: H-6\rightarrow L+1\rangle+c.c.,$
and $\vert H-1\rightarrow L+1;\: H\rightarrow L+2\rangle+c.c.$, respectively.
For decacene, the states involved are $11{}^{3}B_{1g}^{-}$ and $14{}^{3}A_{g}^{-}$,
with their wave functions dominated by the double excitations $\vert H\rightarrow L+1;\, H-1\rightarrow L+5\rangle+c.c.$,
and single excitations $\vert H-1\rightarrow L+8\rangle+c.c.$, respectively. 
\item With the the standard parameters, peak VII for octacene corrresponds
to an $x$-polarized transition to the state $7{}^{3}B{}_{1g}^{-}$
with the double excitations $\vert H\rightarrow L;\, H-2\rightarrow L+3\rangle+c.c.$
dominating its wave function. However, for nonacene and decacene this
peak corresponds to a mixed $x$- and $y$-polarized transition. For
nonacene, the states involved are $7{}^{3}B_{1g}^{-}$ and $11{}^{3}A{}_{g}^{-}$,
with the double excitations $\vert H\rightarrow L+1;\, H-5\rightarrow L\rangle+c.c.$,
and single excitations $\vert H\rightarrow L+5\rangle+c.c.$, contributing
the most to their respective wave functions. For decacene, the states
forming the peak are $8{}^{3}B_{1g}^{-}$ and $11{}^{3}A{}_{g}^{-}$,
of which the former is dominated by the double and triple excitations,
while the latter consists mainly of the single excitations. \\
 \\
 Screened parameter calculations predict peak VII to have a mixed
$x$ and $y$ polarized character for all the oligomers. For the case
of octacene, four states $11{}^{3}B_{1g}^{-}$, $12{}^{3}B_{1g}^{-}$,
$11{}^{3}A{}_{g}^{-}$, and $12{}^{3}A{}_{g}^{-}$, form this peak
and the double excitations dominate the wave function of the first
three states, while $12{}^{3}A{}_{g}^{-}$ is dominated by the single
excitations. For nonacene, two states $11{}^{3}B_{1g}^{-}$ and $12{}^{3}A{}_{g}^{-}$
shape the peak with the triple excitations ($\vert H\rightarrow L;\, H\rightarrow L;\, H-1\rightarrow L+2\rangle+c.c.$)
dominating the wave function of the former, and single excitations
($\vert H-1\rightarrow L+8\rangle+c.c.$) that of the latter. For
decacene, three states, $13{}^{3}B_{1g}^{-},$ $14{}^{3}B_{1g}^{-}$,
and $15{}^{3}A{}_{g}^{-}$, contribute to the peak, with the double
excitations dominating all their wave functions. 
\item In the standard parameter spectrum, peak VIII has a mixed $x$ and
$y$ polarized character for octacene, but only a $y$-polarized character
for nonacene and decacene. For octacene, the peak is formed by states
$8{}^{3}B_{1g}^{-}$ and $12{}^{3}A{}_{g}^{-}$, of which the wave
function of the former is dominated by the triple and double excitations,
while that of the latter by double and single excitations. For nonacene,
the state in question is $13{}^{3}A{}_{g}^{-}$, while for decacene
it is $12{}^{3}A{}_{g}^{-}$, and wave functions in both the cases
are dominated by doubly-excited configurations. \\
 \\
 In the screened parameter calculations, peak VIII exists only for
octacene and nonacene, and is due to an $x$-polarized transition.
For octacene it is formed by the state $13{}^{3}B_{1g}^{-}$, whose
wave function is dominated by both single and double excitations.
For nonacene, the peak is caused by the state $12{}^{3}B_{1g}^{-}$,
whose wave function mainly consists of double excitations. 
\end{enumerate}
\begin{center}
\begin{table}
\caption{Excited states contributing to the triplet absorption spectrum of
octacene computed using the MRSDCI method coupled with the standard
parameters in the PPP model Hamiltonian. The table includes many-particle
dominant contributing configurations, excitation energies, dipole
matrix elements, and relative oscillator strengths (ROS) of various
states. The excitation energies are with respect to the $1^{1}A_{g}^{-}$,
ground state, while the dipole matrix elements, and the ROS, are with
respect to the $1^{3}B_{2u}^{+}$ state. Below, `$+c.c.$' indicates
that the coefficient of charge conjugate of a given configuration
has the same sign, while `$-c.c.$' implies that the two coefficients
have opposite signs.\label{tab:acene-8-pa}}

\centering{}%
\begin{tabular}{cccccc}
\hline 
Peak  & State  & E (eV)  & Transition  & ROS  & dominant contributing configurations\tabularnewline
\hline 
 &  &  & Dipole (\AA{})  &  & \tabularnewline
\hline 
I  & $1^{3}B_{1g}^{-}$  & 3.04  & 2.950  & 0.839  & $|H\rightarrow L+1\left\rangle \right.-c.c.(0.5638)$\tabularnewline
 &  &  &  &  & $|H-1\rightarrow L+2\left\rangle \right.+c.c.(0.1836)$\tabularnewline
II  & $1^{3}A{}_{g}^{-}$  & 3.89  & 0.997  & 0.123  & $|H\rightarrow L;H-1\rightarrow L+1\left\rangle \right.(0.8187)$\tabularnewline
 &  &  &  &  & $|H\rightarrow L;H-2\rightarrow L+2\left\rangle \right.(0.1135)$\tabularnewline
III  & $3{}^{3}A{}_{g}^{-}$  & 4.28  & 0.355  & 0.017  & $|H-1\rightarrow L+3\left\rangle \right.-c.c.(0.3698)$\tabularnewline
 &  &  &  &  & $|H\rightarrow L+5\left\rangle \right.-c.c.(0.3435)$\tabularnewline
 & $2^{3}B_{1g}^{-}$  & 4.35  & 0.579  & 0.046  & $|H\rightarrow L+4\left\rangle \right.-c.c.(0.5362)$\tabularnewline
 &  &  &  &  & $|H-1\rightarrow L+2\left\rangle \right.+c.c.(0.1999)$\tabularnewline
IV  & $3^{3}B_{1g}^{-}$  & 5.06  & 2.494  & 1.000  & $|H\rightarrow L;H\rightarrow L+3\left\rangle \right.-c.c.(0.4566)$\tabularnewline
 &  &  &  &  & $|H-1\rightarrow L+2\left\rangle \right.+c.c.(0.3011)$\tabularnewline
V  & $4^{3}B_{1g}^{-}$  & 5.26  & 1.199  & 0.240  & $|H-1\rightarrow L+2\left\rangle \right.+c.c.(0.3688)$\tabularnewline
 &  &  &  &  & $|H\rightarrow L;H\rightarrow L+3\left\rangle \right.-c.c.(0.2757)$\tabularnewline
VI  & $9^{3}A{}_{g}^{-}$  & 5.82  & 0.588  & 0.064  & $|H\rightarrow L+5\left\rangle \right.-c.c.(0.3967)$\tabularnewline
 &  &  &  &  & $|H-3\rightarrow L+1\left\rangle \right.-c.c.(0.3809)$\tabularnewline
VII  & $7^{3}B_{1g}^{-}$  & 6.10  & 1.055  & 0.216  & $|H\rightarrow L;H-2\rightarrow L+3\left\rangle \right.-c.c.(0.4326)$\tabularnewline
 &  &  &  &  & $|H\rightarrow L+1;H-1\rightarrow L+3\left\rangle \right.-c.c.(0.1274)$\tabularnewline
VIII  & $8^{3}B_{1g}^{-}$  & 6.29  & 0.864  & 0.149  & $|H\rightarrow L+1;H\rightarrow L+1;H-1\rightarrow L\left\rangle \right.-c.c.(0.4030)$\tabularnewline
 &  &  &  &  & $|H\rightarrow L;H-2\rightarrow L+3\left\rangle \right.-c.c.(0.3504)$\tabularnewline
 & $12^{3}A{}_{g}^{-}$  & 6.29  & 0.149  & 0.004  & $|H-1\rightarrow L+1;H\rightarrow L+2\left\rangle \right.-c.c.(0.3707)$\tabularnewline
 &  &  &  &  & $|H-3\rightarrow L+4\left\rangle \right.-c.c.(0.2968)$\tabularnewline
\hline 
\end{tabular}
\end{table}

\par\end{center}

\begin{center}
\begin{table}
\caption{Excited states contributing to the triplet absorption spectrum of
octacene computed using the MRSDCI method coupled with the screened
parameters in the PPP model Hamiltonian. The rest of the information
is same as that in Table \ref{tab:acene-8-pa}.\label{tab:acene-8-pa-scr}}

\centering{}%
\begin{tabular}{cccccc}
\hline 
Peak  & State  & E (eV)  & Transition  & ROS  & dominant contributing configurations\tabularnewline
\hline 
 &  &  & Dipole (\AA{})  &  & \tabularnewline
\hline 
I  & $1^{3}B_{1g}^{-}$  & 1.97  & 4.639  & 1.000  & $|H\rightarrow L+1\left\rangle \right.+c.c.(0.5936)$\tabularnewline
 &  &  &  &  & $|H-1\rightarrow L+2\left\rangle \right.+c.c.(0.0990)$\tabularnewline
II  & $1^{3}A{}_{g}^{-}$  & 2.92  & 0.996  & 0.064  & $|H\rightarrow L;H-1\rightarrow L+1\left\rangle \right.(0.8333)$\tabularnewline
 &  &  &  &  & $|H-1\rightarrow L;H-1\rightarrow L+2\left\rangle \right.-c.c.(0.1090)$\tabularnewline
III  & $3{}^{3}B_{1g}^{-}$  & 3.39  & 0.465  & 0.017  & $|H\rightarrow L+3\left\rangle \right.+c.c.(0.4280)$\tabularnewline
 &  &  &  &  & $|H-1\rightarrow L+2\left\rangle \right.+c.c.(0.3614)$\tabularnewline
IV  & $3{}^{3}A{}_{g}^{-}$  & 3.68  & 0.621  & 0.034  & $|H\rightarrow L+5\left\rangle \right.+c.c.(0.4915)$\tabularnewline
 &  &  &  &  & $|H-1\rightarrow L+4\left\rangle \right.+c.c.(0.3220)$\tabularnewline
V  & $4{}^{3}B_{1g}^{-}$  & 4.01  & 2.601  & 0.641  & $|H\rightarrow L;H\rightarrow L+4\left\rangle \right.-c.c.(0.5605)$\tabularnewline
 &  &  &  &  & $|H\rightarrow L+1;H-1\rightarrow L+4\left\rangle \right.-c.c.(0.1764)$\tabularnewline
VI  & $7{}^{3}B_{1g}^{-}$  & 4.67  & 1.017  & 0.114  & $|H\rightarrow L+1;H-5\rightarrow L\left\rangle \right.-c.c.(0.5190)$\tabularnewline
 &  &  &  &  & $|H\rightarrow L+1;H-1\rightarrow L+4\left\rangle \right.-c.c.(0.2712)$\tabularnewline
VII  & $11{}^{3}A{}_{g}^{-}$  & 5.13  & 0.151  & 0.003  & $|H-1\rightarrow L+1;H\rightarrow L+2\left\rangle \right.-c.c.(0.3311)$\tabularnewline
 &  &  &  &  & $|H-1\rightarrow L+8\left\rangle \right.-c.c.(0.2700)$\tabularnewline
 & $11{}^{3}B_{1g}^{-}$  & 5.15  & 0.768  & 0.072  & $|H\rightarrow L+1;H-1\rightarrow L+4\left\rangle \right.-c.c.(0.4913)$\tabularnewline
 &  &  &  &  & $|H\rightarrow L+1;H-5\rightarrow L\left\rangle \right.-c.c.(0.2874)$\tabularnewline
 & $12{}^{3}B_{1g}^{-}$  & 5.18  & 0.345  & 0.015  & $|H\rightarrow L+1;H-5\rightarrow L\left\rangle \right.-c.c.(0.4987)$\tabularnewline
 &  &  &  &  & $|H\rightarrow L+1;H-1\rightarrow L+4\left\rangle \right.-c.c.(0.2693)$\tabularnewline
 & $12{}^{3}A{}_{g}^{-}$  & 5.19  & 0.148  & 0.003  & $|H-1\rightarrow L+8\left\rangle \right.+c.c.(0.3959)$\tabularnewline
 &  &  &  &  & $|H\rightarrow L;H-1\rightarrow L+3\left\rangle \right.-c.c.(0.2687)$\tabularnewline
VIII  & $13{}^{3}B_{1g}^{-}$  & 5.39  & 0.777  & 0.077  & $|H-1\rightarrow L+8\left\rangle \right.+c.c.(0.3959)$\tabularnewline
 &  &  &  &  & $|H\rightarrow L;H-1\rightarrow L+3\left\rangle \right.-c.c.(0.2687)$\tabularnewline
\hline 
\end{tabular}
\end{table}

\par\end{center}

\begin{center}
\begin{table}
\caption{Excited states contributing to the triplet absorption spectrum of
nonacene computed using the MRSDCI method coupled with the standard
parameters in the PPP model Hamiltonian. The rest of the information
is same as that in Table \ref{tab:acene-8-pa}.\label{tab:acene-9-pa}}

\centering{}%
\begin{tabular}{cccccc}
\hline 
Peak  & State  & E (eV)  & Transition  & ROS  & dominant contributing configurations\tabularnewline
\hline 
 &  &  & Dipole (\AA{})  &  & \tabularnewline
\hline 
I  & $1^{3}B_{1g}^{-}$  & 2.56  & 3.331  & 0.785  & $|H\rightarrow L+1\left\rangle \right.+c.c.(0.5522)$\tabularnewline
 &  &  &  &  & $|H-1\rightarrow L+2\left\rangle \right.+c.c.(0.1825)$\tabularnewline
II  & $1^{3}A{}_{g}^{-}$  & 3.30  & 1.018  & 0.095  & $|H\rightarrow L;H-1\rightarrow L+1\left\rangle \right.(0.8148)$\tabularnewline
 &  &  &  &  & $|H\rightarrow L;H-1\rightarrow L+3\left\rangle \right.c.c.(0.1109)$\tabularnewline
III  & $2^{3}B_{1g}^{-}$  & 3.79  & 0.686  & 0.049  & $|H\rightarrow L+3\left\rangle \right.-c.c.(0.5156)$\tabularnewline
 &  &  &  &  & $|H-1\rightarrow L+2\left\rangle \right.+c.c.(0.2173)$\tabularnewline
 & $3{}^{3}A{}_{g}^{-}$  & 3.89  & 0.352  & 0.013  & $|H\rightarrow L;H-1\rightarrow L+1\left\rangle \right.(0.4041)$\tabularnewline
 &  &  &  &  & $|H-1\rightarrow L+4\left\rangle \right.-c.c.(0.3279)$\tabularnewline
IV  & $3{}^{3}B_{1g}^{-}$  & 4.39  & 0.790  & 0.076  & $|H-1\rightarrow L+2\left\rangle \right.+c.c.(0.4226)$\tabularnewline
 &  &  &  &  & $|H\rightarrow L;H\rightarrow L;H-1\rightarrow L+2\left\rangle \right.+c.c.(0.2222)$\tabularnewline
V  & $4^{3}B_{1g}^{-}$  & 4.66  & 2.789  & 1.000  & $|H\rightarrow L;H\rightarrow L+4\left\rangle \right.+c.c.(0.5047)$\tabularnewline
 &  &  &  &  & $|H-1\rightarrow L;H\rightarrow L+5\left\rangle \right.-c.c.(0.2034)$\tabularnewline
VI  & $6^{3}B_{1g}^{-}$  & 5.14  & 0.418  & 0.025  & $|H\rightarrow L+1;H\rightarrow L+1;H-1\rightarrow L\left\rangle \right.+c.c.(0.4111)$\tabularnewline
 &  &  &  &  & $|H\rightarrow L;H\rightarrow L+4\left\rangle \right.+c.c.(0.2905)$\tabularnewline
VII  & $11{}^{3}A{}_{g}^{-}$  & 5.39  & 0.517  & 0.040  & $|H\rightarrow L+5\left\rangle \right.-c.c.(0.3654)$\tabularnewline
 &  &  &  &  & $|H-1\rightarrow L+4\left\rangle \right.-c.c.(0.3219)$\tabularnewline
 & $7^{3}B_{1g}^{-}$  & 5.44  & 1.438  & 0.310  & $|H\rightarrow L+1;H-5\rightarrow L\left\rangle \right.-c.c.(0.4435)$\tabularnewline
 &  &  &  &  & $|H\rightarrow L+1;H\rightarrow L+4\left\rangle \right.-c.c.(0.3271)$\tabularnewline
VIII  & $13{}^{3}A{}_{g}^{-}$  & 5.75  & 0.428  & 0.029  & $|H-1\rightarrow L+1;H\rightarrow L+2\left\rangle \right.-c.c.(0.2992)$\tabularnewline
 &  &  &  &  & $|H\rightarrow L;H-1\rightarrow L+3\left\rangle \right.-c.c.(0.2305)$\tabularnewline
\hline 
\end{tabular}
\end{table}

\par\end{center}

\begin{center}
\begin{table}
\caption{Excited states contributing to the triplet absorption spectrum of
nonacene computed using the MRSDCI method coupled with the screened
parameters in the PPP model Hamiltonian. The rest of the information
is same as that in Table \ref{tab:acene-8-pa}.\label{tab:acene-9-pa-scr}}

\centering{}%
\begin{tabular}{cccccc}
\hline 
Peak  & State  & E (eV)  & Transition  & ROS  & dominant contributing configurations\tabularnewline
\hline 
 &  &  & Dipole (\AA{})  &  & \tabularnewline
\hline 
I  & $1^{3}B_{1g}^{-}$  & 1.86  & 5.218  & 1.000  & $|H\rightarrow L+1\left\rangle \right.+c.c.(0.5919)$\tabularnewline
 &  &  &  &  & $|H-1\rightarrow L+2\left\rangle \right.-c.c.(0.1143)$\tabularnewline
II  & $1^{3}A{}_{g}^{-}$  & 2.61  & 1.032  & 0.055  & $|H\rightarrow L;H-1\rightarrow L+1\left\rangle \right.(0.8266)$\tabularnewline
 &  &  &  &  & $|H-1\rightarrow L;H\rightarrow L+2\left\rangle \right.+c.c.(0.1122)$\tabularnewline
III  & $3^{3}B_{1g}^{-}$  & 3.14  & 0.711  & 0.031  & $|H\rightarrow L+3\left\rangle \right.+c.c.(0.4278)$\tabularnewline
 &  &  &  &  & $|H-1\rightarrow L+2\left\rangle \right.-c.c.(0.3551)$\tabularnewline
IV  & $3^{3}A{}_{g}^{-}$  & 3.58  & 0.514  & 0.019  & $|H\rightarrow L+6\left\rangle \right.+c.c.(0.4657)$\tabularnewline
 &  &  &  &  & $|H-1\rightarrow L+4\left\rangle \right.-c.c.(0.3453)$\tabularnewline
V  & $6^{3}B_{1g}^{-}$  & 4.09  & 2.577  & 0.535  & $|H\rightarrow L;H\rightarrow L+4\left\rangle \right.+c.c.(0.5088)$\tabularnewline
 &  &  &  &  & $|H-1\rightarrow L;H\rightarrow L+1;H\rightarrow L+1\left\rangle \right.+c.c.(0.2302)$\tabularnewline
VI  & $10^{3}A{}_{g}^{-}$  & 4.52  & 0.324  & 0.009  & $|H-1\rightarrow L+1;H\rightarrow L+2\left\rangle \right.-c.c.(0.3581)$\tabularnewline
 &  &  &  &  & $|H\rightarrow L;H-1\rightarrow L+3\left\rangle \right.-c.c.(0.2788)$\tabularnewline
 & $8^{3}B_{1g}^{-}$  & 4.53  & 1.434  & 0.184  & $|H\rightarrow L+1;H-6\rightarrow L+1\left\rangle \right.-c.c.(0.5621)$\tabularnewline
 &  &  &  &  & $|H-2\rightarrow L+3\left\rangle \right.-c.c.(0.1869)$\tabularnewline
VII  & $11^{3}B_{1g}^{-}$  & 4.87  & 0.398  & 0.015  & $|H\rightarrow L;H\rightarrow L;H-1\rightarrow L+2\left\rangle \right.-c.c.(0.4361)$\tabularnewline
 &  &  &  &  & $|H-1\rightarrow L;H-1\rightarrow L;H\rightarrow L+1\left\rangle \right.+c.c.(0.3273)$\tabularnewline
 & $12^{3}A{}_{g}^{-}$  & 4.92  & 0.215  & 0.004  & $|H-1\rightarrow L+8\left\rangle \right.+c.c.(0.4643)$\tabularnewline
 &  &  &  &  & $|H\rightarrow L;H-1\rightarrow L+3\left\rangle \right.-c.c.(0.2204)$\tabularnewline
VIII  & $12^{3}B_{1g}^{-}$  & 5.14  & 0.884  & 0.079  & $|H\rightarrow L+1;H-6\rightarrow L+1\left\rangle \right.-c.c.(0.5873)$\tabularnewline
 &  &  &  &  & $|H\rightarrow L;H-1\rightarrow L+2\left\rangle \right.-c.c.(0.0756)$\tabularnewline
\hline 
\end{tabular}
\end{table}

\par\end{center}

\begin{center}
\begin{table}
\caption{Excited states contributing to the triplet absorption spectrum of
decacene computed using the MRSDCI method coupled with the standard
parameters in the PPP model Hamiltonian. The rest of the information
is same as that in Table \ref{tab:acene-8-pa}.\label{tab:acene-10-pa}}

\centering{}%
\begin{tabular}{cccccc}
\hline 
Peak  & State  & E (eV)  & Transition  & ROS  & dominant contributing configurations\tabularnewline
\hline 
 &  &  & Dipole (\AA{})  &  & \tabularnewline
\hline 
I  & $1^{3}B_{1g}^{-}$  & 2.49  & 3.615  & 1.000  & $|H\rightarrow L+1\left\rangle \right.-c.c.(0.5461)$\tabularnewline
 &  &  &  &  & $|H-1\rightarrow L+2\left\rangle \right.+c.c.(0.1918)$\tabularnewline
II  & $1^{3}A{}_{g}^{-}$  & 3.16  & 1.097  & 0.117  & $|H\rightarrow L;H-1\rightarrow L+1\left\rangle \right.(0.8189)$\tabularnewline
 &  &  &  &  & $|H\rightarrow L;H-1\rightarrow L+3\left\rangle \right.-c.c.(0.1177)$\tabularnewline
III  & $2^{3}B_{1g}^{-}$  & 3.62  & 0.800  & 0.071  & $|H\rightarrow L+3\left\rangle \right.-c.c.(0.5228)$\tabularnewline
 &  &  &  &  & $|H-1\rightarrow L+2\left\rangle \right.+c.c.(0.1965)$\tabularnewline
IV  & $3^{3}B_{1g}^{-}$  & 4.35  & 0.793  & 0.084  & $|H-1\rightarrow L+2\left\rangle \right.+c.c.(0.4745)$\tabularnewline
 &  &  &  &  & $|H-1\rightarrow L+4\left\rangle \right.-c.c.(0.1760)$\tabularnewline
V  & $4^{3}B_{1g}^{-}$  & 4.62  & 2.534  & 0.912  & $|H\rightarrow L;H\rightarrow L+4\left\rangle \right.-c.c.(0.4009)$\tabularnewline
 &  &  &  &  & $|H\rightarrow L+7\left\rangle \right.-c.c.(0.2788)$\tabularnewline
VI  & $6^{3}B_{1g}^{-}$  & 5.22  & 1.183  & 0.225  & $|H\rightarrow L+1;H\rightarrow L+1;H-1\rightarrow L\left\rangle \right.-c.c.(0.3604)$\tabularnewline
 &  &  &  &  & $|H\rightarrow L+1;H-6\rightarrow L\left\rangle \right.-c.c.(0.3314)$\tabularnewline
VII  & $11^{3}A{}_{g}^{-}$  & 5.42  & 0.493  & 0.040  & $|H\rightarrow L+6\left\rangle \right.+c.c.(0.3537)$\tabularnewline
 &  &  &  &  & $|H-1\rightarrow L+4\left\rangle \right.-c.c.(0.3086)$\tabularnewline
 & $8^{3}B_{1g}^{-}$  & 5.44  & 1.179  & 0.232  & $|H\rightarrow L+1;H-1\rightarrow L;H-1\rightarrow L\left\rangle \right.-c.c.(0.3928)$\tabularnewline
 &  &  &  &  & $|H\rightarrow L+1;H-6\rightarrow L\left\rangle \right.-c.c.(0.3365)$\tabularnewline
VIII  & $12^{3}A{}_{g}^{-}$  & 5.70  & 0.456  & 0.036  & $|H-1\rightarrow L+1;H\rightarrow L+2\left\rangle \right.-c.c.(0.3473)$\tabularnewline
 &  &  &  &  & $|H\rightarrow L+1;H\rightarrow L+3\left\rangle \right.-c.c.(0.2317)$\tabularnewline
\hline 
\end{tabular}
\end{table}

\par\end{center}

\begin{center}
\begin{table}
\caption{Excited states contributing to the triplet absorption spectrum of
decacene computed using the MRSDCI method coupled with the screened
parameters in the PPP model Hamiltonian. The rest of the information
is same as that in Table \ref{tab:acene-8-pa}.\label{tab:acene-10-pa-scr}}

\centering{}%
\begin{tabular}{cccccc}
\hline 
Peak  & State  & E (eV)  & Transition  & ROS  & dominant contributing configurations\tabularnewline
\hline 
 &  &  & Dipole (\AA{})  &  & \tabularnewline
\hline 
I  & $1^{3}B_{1g}^{-}$  & 1.72  & 5.841  & 1.000  & $|H\rightarrow L+1\left\rangle \right.-c.c.(0.5872)$\tabularnewline
 &  &  &  &  & $|H-1\rightarrow L+2\left\rangle \right.-c.c.(0.1112)$\tabularnewline
II  & $1^{3}A{}_{g}^{-}$  & 2.33  & 1.074  & 0.046  & $|H\rightarrow L;H-1\rightarrow L+1\left\rangle \right.(0.8198)$\tabularnewline
 &  &  &  &  & $|H\rightarrow L;H-1\rightarrow L+2\left\rangle \right.-c.c.(0.0986)$\tabularnewline
III  & $3^{3}B_{1g}^{-}$  & 2.89  & 0.831  & 0.034  & $|H\rightarrow L+3\left\rangle \right.-c.c.(0.4352)$\tabularnewline
 &  &  &  &  & $|H-1\rightarrow L+2\left\rangle \right.-c.c.(0.3443)$\tabularnewline
IV  & $6^{3}B_{1g}^{-}$  & 3.80  & 2.695  & 0.470  & $|H\rightarrow L;H\rightarrow L+5\left\rangle \right.-c.c.(0.5362)$\tabularnewline
 &  &  &  &  & $|H\rightarrow L+1;H-1\rightarrow L+5\left\rangle \right.-c.c.(0.1922)$\tabularnewline
V  & $8{}^{3}B_{1g}^{-}$  & 4.24  & 1.266  & 0.116  & $|H-1\rightarrow L;H\rightarrow L+6\left\rangle \right.-c.c.(0.4998)$\tabularnewline
 &  &  &  &  & $|H\rightarrow L+1;H-1\rightarrow L+5\left\rangle \right.-c.c.(0.3005)$\tabularnewline
VI  & $11{}^{3}B_{1g}^{-}$  & 4.70  & 0.791  & 0.050  & $|H\rightarrow L+1;H-1\rightarrow L+5\left\rangle \right.-c.c.(0.4721)$\tabularnewline
 &  &  &  &  & $|H\rightarrow L+1;H-6\rightarrow L\left\rangle \right.-c.c.(0.3044)$\tabularnewline
 & $14^{3}A{}_{g}^{-}$  & 4.72  & 0.135  & 0.001  & $|H-1\rightarrow L+8\left\rangle \right.-c.c.(0.4705)$\tabularnewline
 &  &  &  &  & $|H-3\rightarrow L+5\left\rangle \right.-c.c.(0.2190)$\tabularnewline
VII  & $13{}^{3}B_{1g}^{-}$  & 4.90  & 1.001  & 0.084  & $|H\rightarrow L+1;H-5\rightarrow L+1\left\rangle \right.-c.c.(0.3314)$\tabularnewline
 &  &  &  &  & $|H\rightarrow L+13\left\rangle \right.-c.c.(0.3095)$\tabularnewline
 & $14{}^{3}B_{1g}^{-}$  & 4.91  & 0.638  & 0.034  & $|H\rightarrow L+13\left\rangle \right.-c.c.(0.4517)$\tabularnewline
 &  &  &  &  & $|H\rightarrow L+1;H-5\rightarrow L+1\left\rangle \right.-c.c.(0.2219)$\tabularnewline
 & $15^{3}A{}_{g}^{-}$  & 4.91  & 0.162  & 0.002  & $|H-1\rightarrow L+1;H\rightarrow L+4\left\rangle \right.-c.c.(0.4666)$\tabularnewline
 &  &  &  &  & $|H-1\rightarrow L+1;H-2\rightarrow L+2\left\rangle \right.(0.4449)$\tabularnewline
\hline 
\end{tabular}
\end{table}

\par\end{center}

\end{suppinfo}

\bibliography{long_acene}
\begin{tocentry}

\includegraphics[scale=0.3]{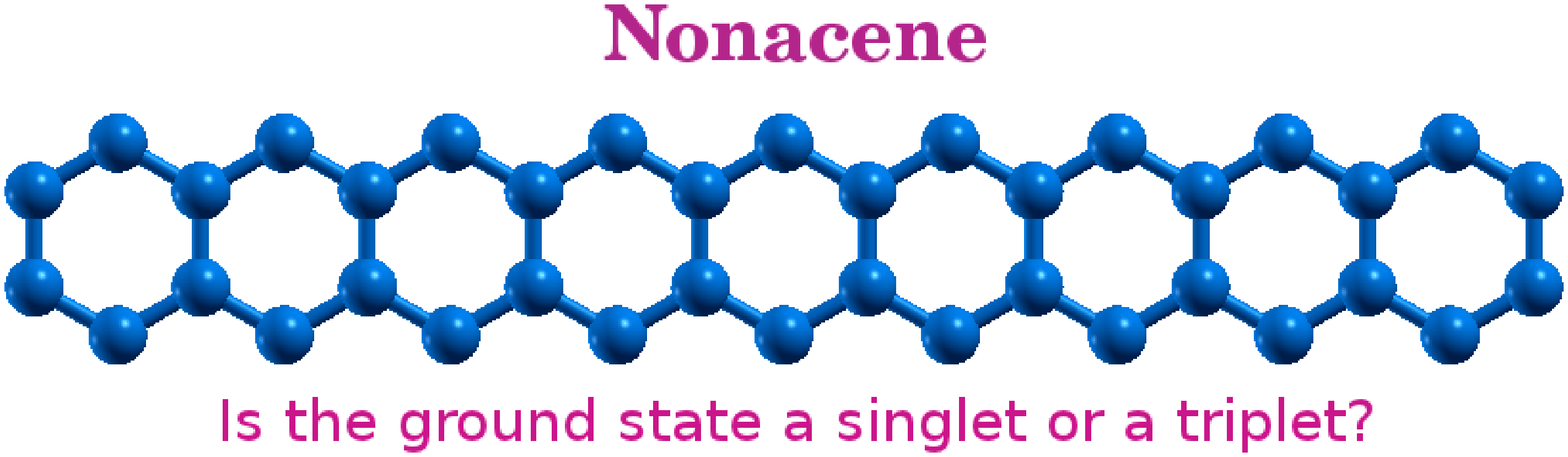}

\end{tocentry}
\end{document}